\documentclass[aps,prb,reprint,showpacs,superscriptaddress,groupedaddress]{revtex4-1}
\usepackage{graphicx}
\usepackage{amsmath}
\usepackage{amssymb}
\usepackage{dcolumn}
\usepackage{latexsym}
\usepackage{rotating}
\usepackage[usenames,dvipsnames]{xcolor}
\usepackage{float}
\usepackage{epsfig}
\usepackage{psfrag}
\usepackage{natbib}
\usepackage{bm}
\usepackage{eucal}
\usepackage{braket}
\usepackage{enumerate}
\usepackage{longtable}
\usepackage{subfigure}
\usepackage{bm}
\usepackage{hyperref}
\usepackage{amsfonts}
\usepackage{dcolumn}
\usepackage{bm}
\usepackage{subfigure}

\newcommand{\be}{\begin{equation}}
\newcommand{\ee}{\end{equation}}
\newcommand{\bn}{\begin{eqnarray}}
\newcommand{\en}{\end{eqnarray}}

\usepackage{color} 

\usepackage{hyperref}
\begin{document}

\author{S. Koley$^{1}$}
\thanks{Present address: Department of Physics, Amity Institute of Applied Sciences, Kolkata, 700135, India.}
\author{S. Basu$^{2}$}

\title{Intercalated phosphorene for improved spintronic applications}
\affiliation{$^{1}$Department of Physics, North Eastern Hill University,
Shillong, Meghalaya, 793022 India.\\
$^{2}$Department of Physics, Indian Institute of Technology Guwahati, Assam, 781039, India
}

\begin{abstract}
\noindent
In this work we study the intercalation of monolayer phosphorene with nitrogen, 
lithium and calcium for exploring prospects of spintronic applications. The electronic and the magnetic properties of the intercalated structure are 
investigated via density functional theory to obtain the band structure and the 
spin polarized density of states. Albeit the band structure data show vanishing
 band gap, a noticeable difference emerges in the densities of the up and the 
down spin states induced by the intercalants. To evaluate the performance of the
 intercalated phosphorene, the spintronic order parameter, measuring the 
asymmetry among the up and the down spin densities of states, is computed which 
clearly shows evolution of improved spintronic properties at large intercalant 
densities. Further, larger atomic numbers of the intercalants seem to aid the 
performance of phosphorene as a spintronic material.
\end{abstract}
\maketitle
\section{Introduction}
Modern civilization is based on integrated circuits (IC) made of silicon chips.
According to Moore's Law,
the efficiency of transistors in IC will double every two years.
However, due to the limit of the transistor size that could fit on
silicon chips, the semiconductor industry continuously tries to shrink the sizes of the electronic components on ICs, thereby enhancing the performance of the 
computers. Further, to overcome the hurdle related to the miniaturization of 
the components 
new methods and novel materials are being discovered which can dramatically 
transform the electronics industry. Thus search of new suitable materials 
that can replace silicon in IC constitutes new motivation for materials research\cite{felser}. 
The quest for new and non-traditional materials whose properties can be 
controlled easily 
and the spins can be made to carry information instead of the charges, has 
contributed 
to the development of a field called spintronics. Here the spins make the 
electrons act like tiny magnets where the
'up' and the 'down' states behave similar to the positive and the negative charges which can store information in a binary format \cite{baibich}. 
The necessary properties of these spintronic materials are tunable bandgap in both the up-spin and 
the down-spin channels and large magnetization values that are computed using $m=(n_{\uparrow}-n_{\downarrow})\mu_B$, 
$m$ being the total magnetization and $n_{\uparrow}$ and $n_{\downarrow}$ 
are the average occupation densities for the spin-up and spin-down states, respectively of the carriers\cite{sun}. 
So the magnetic semiconductors with relatively large bandgap between the up-spin and the 
down-spin channels fall into this category.  Intercalated phosphorene is 
possibly one such material and hence is studied here.

A lot of bulk and monolayer semiconductors like bismuthene, GaAs, and ZnO have 
large bandgaps and doping induced magnetic properties can also be achieved, but nonetheless, the values of
magnetization remain insignificant for any practical applications\cite{JAP,natmet}. Recent interest into monolayer 
materials like graphene, MoS$_2$, silicene, borophene, and phosphorene have drawn attention in the 
search of spintronic materials. Among them graphene has no (or vanishingly small) band gap, however 
pristine phosphorene and MoS$_{2}$ have bandgaps\cite{ding,mak,tran}. Not only these 
monolayer materials, study of spintronics include a diverse area of ferromagnetic, 
antiferromagnetic, half-metallic, topological insulators, magnetic 
semiconductors and spin gapless semiconductors\cite{choudhuri}. Conceptualizing spintronics started 
with another important discovery of giant magnetoresistance(GMR)\cite{baibich,binaschprb} where the GMR 
effect is employed to make two-state memory cells and switches. Inspite of the 
advantages, construction of spintronics materials has a great deal of challenges, such as, having
 fully spin polarized materials, the manipulation of spin orientation of the carriers, 
spin transport and modeling of spin injected devices, mismatch of Fermi energies at the junction of two dissimilar
materials etc.

To solve these issues, a lot of materials have been devised with  
distinct electronic and magnetic properties of matter, monoloayer materials are 
one among them. Since the discovery of graphene, attention is drawn towards other 
materials with same structure leading to renewed interest in phosphorene: 
monolayer black phosphorus\cite{akhtar}. Experimental and theoretical study of phosphorene 
show a direct band gap and high carrier mobility. However the nature of the band gap is not beyond debate, where
the valence band maximum is theoretically found to be shifted from the zone center \cite{Wang,Rodin}. Further, both the
mobility and the band gap can be controlled via tuning the layer thickness of phosphorene. The band gap in a 
single layer of phosphorene yields a large on-off current ratio ($\sim 10^{5}$) which should be relevant for
device applications\cite{LLi}.

Phosphorene in its stable form contains four P molecules and these four atoms form a tetrahedron with each
P atom having three neighbour. All the P$_4$ units are joined to set up 
continuous layers. Comparing the structure with graphene $y$-axis is named as the zigzag direction and $x$-axis as the armchair direction.
Earlier studies have shown that foreign materials can induce magnetization when doped in monolayer 
materials, such as, phosphorene, silicene, borophene and 
bismuthene\cite{yang,kadioglu,mirzaei,hu,jiang,luo,babar,yu,wang,li,a,b,c,d}. In phosphorene, the 3d-transition metal dopants induce strong
magnetic character\cite{d}, particularly Mn doping yields large magnetization. Besides some of the studies show  
the emergence of dilute magnetic semiconducting properties as well upon doping with other elements. Earlier research has also proof
 of magnetization in doping with non-metals, which results in 
semiconductors without a spin-gap. 
Since parent phosphorene has high carrier mobility, it is possible to tune 
the magnetization using intercalation and hence create a better spintronic material. 
\begin{figure}
(a)
\includegraphics[angle=0,width=0.8\columnwidth]{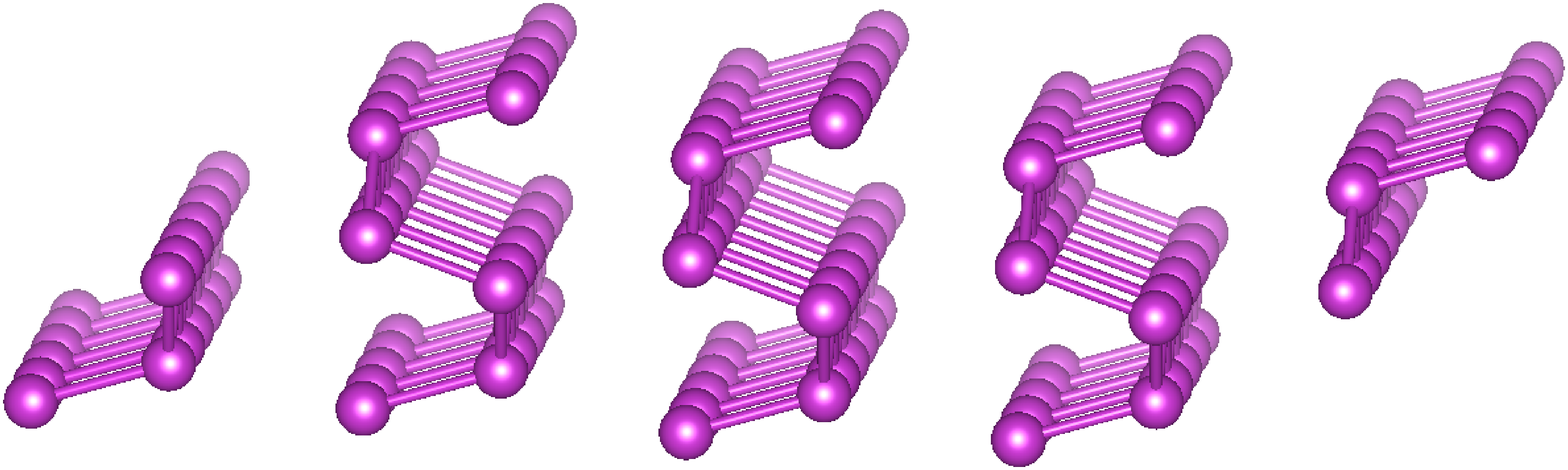}

(b)
\includegraphics[angle=0,width=0.8\columnwidth]{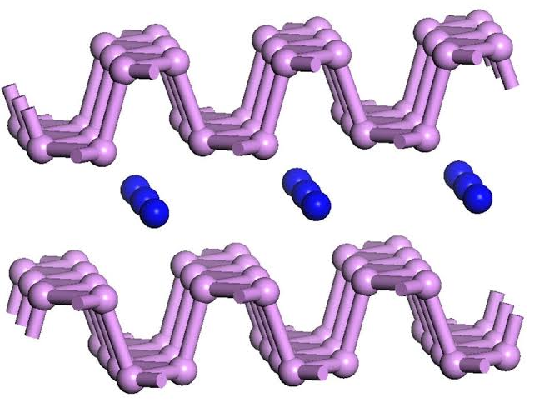}

(c)
\includegraphics[angle=270,width=0.8\columnwidth]{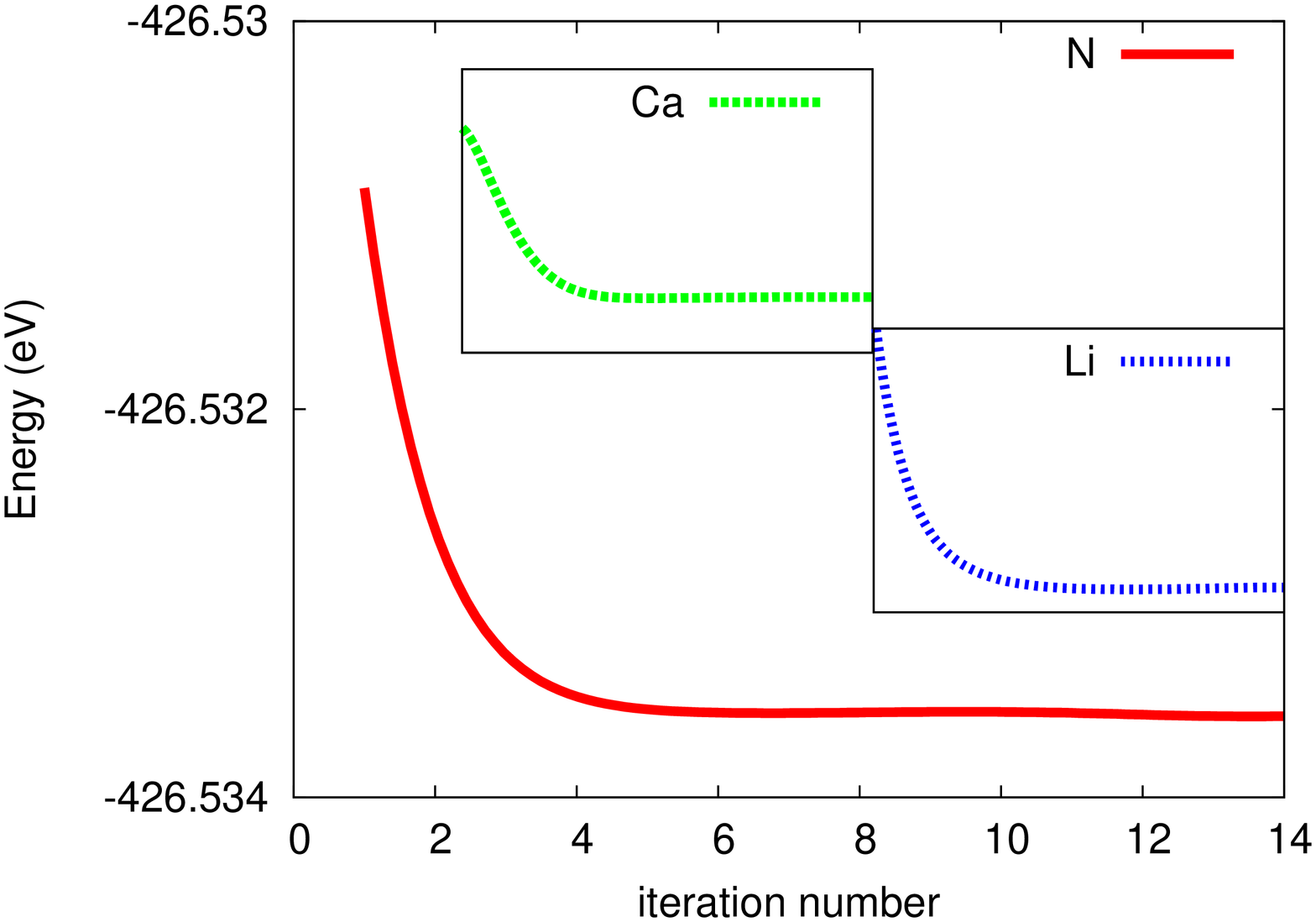}
\caption{(Color Online) Crystal structure of (a) pristine black phosphorene and (b) nitrogen intercalated phosphorene. Blue atoms are the nitrogen atoms and 
the violet atoms are the phosphorus atoms. (c) Plot showing the energy convergence for 
final iterations in the scf method of the density functional theory calculations. N, Li and Ca stand for nitrogen, lithium and calcium respectively.}
\label{fig1}
\end{figure}

\begin{figure*}
(a)
\includegraphics[angle=0,width=0.5\columnwidth]{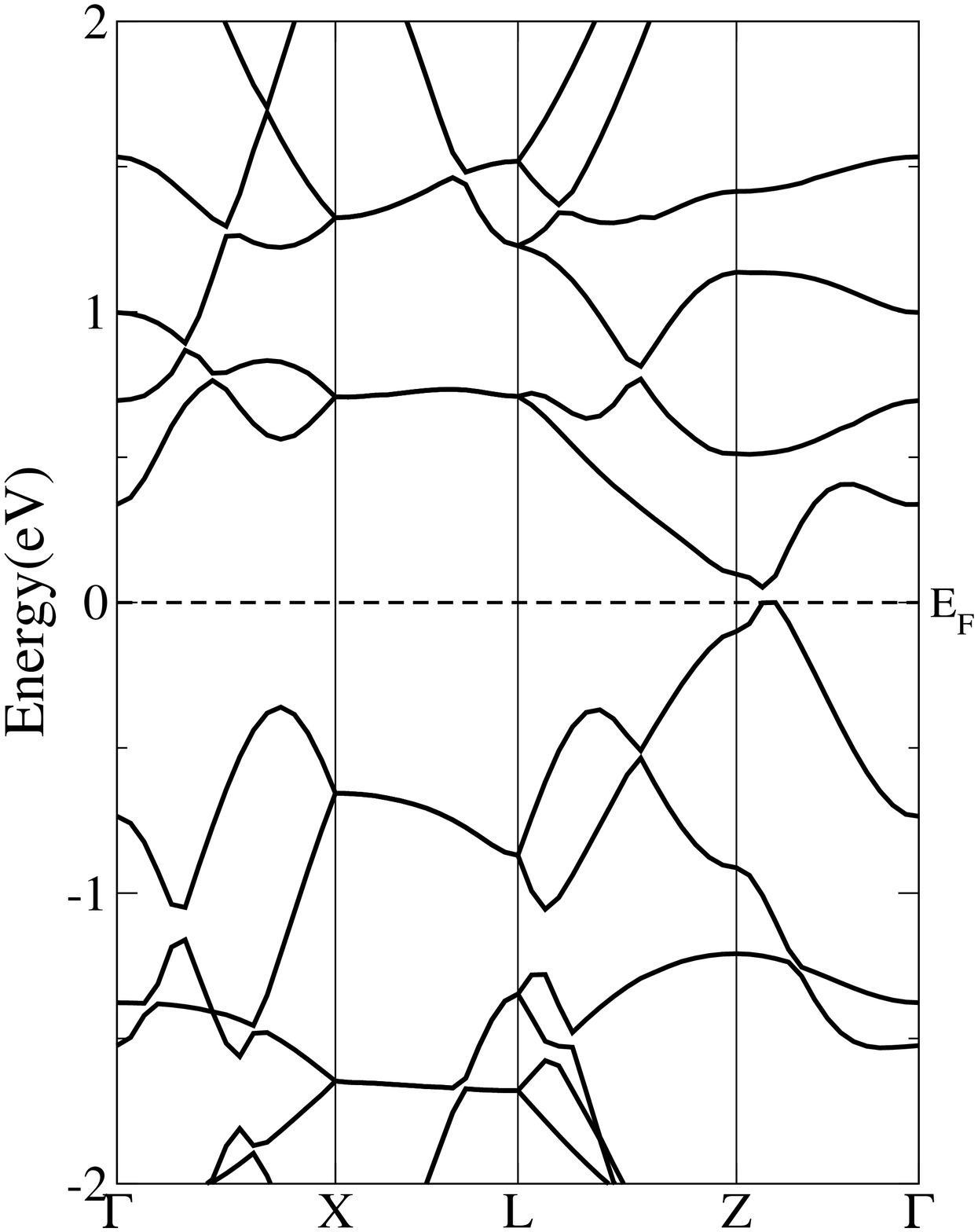}
(b)
\includegraphics[angle=0,width=0.5\columnwidth]{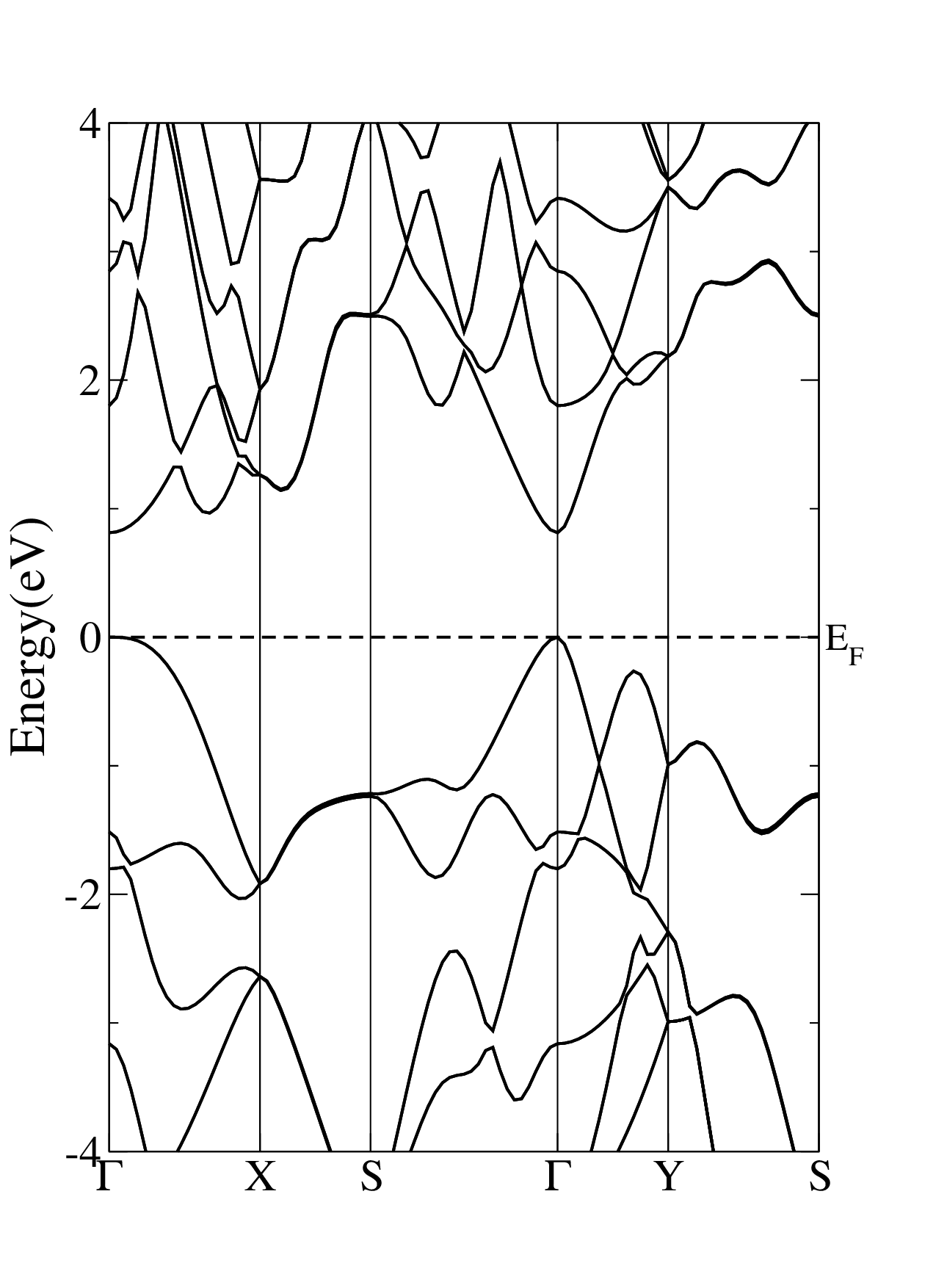}

(c)
\includegraphics[angle=270,width=0.5\columnwidth]{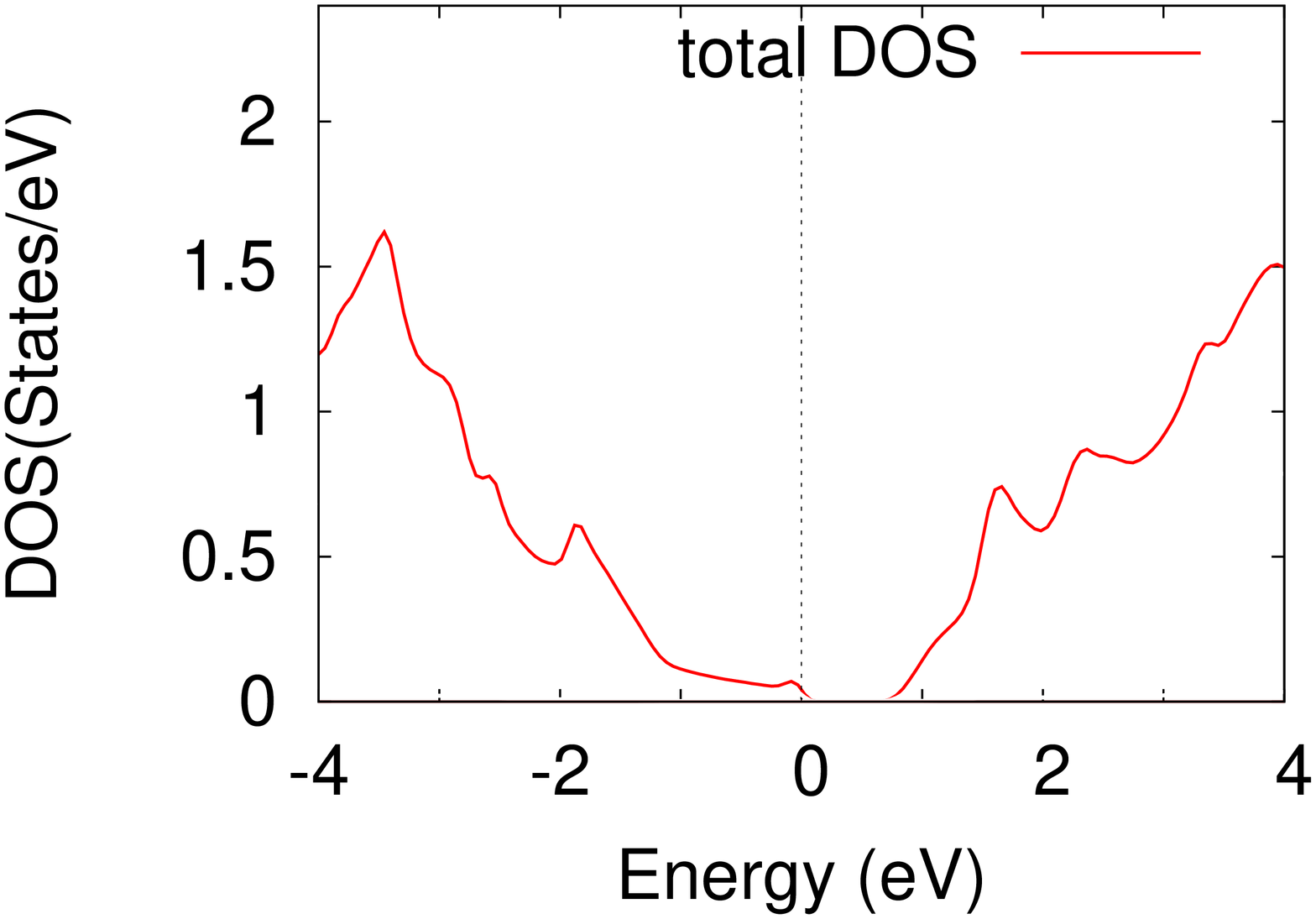}
(d)
\includegraphics[angle=270,width=0.5\columnwidth]{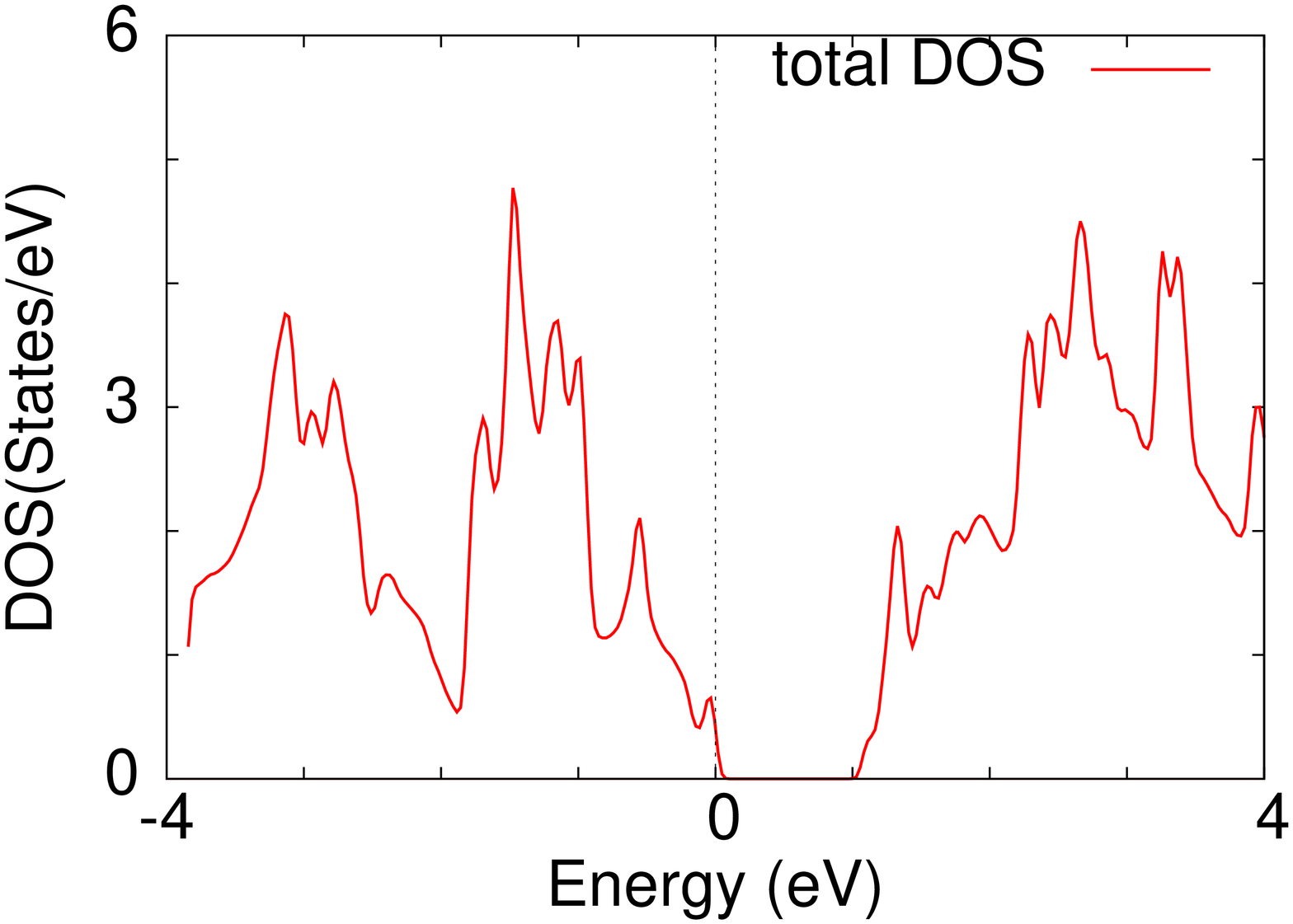}
\caption{(Color Online) (a) and (c) denote the band structure and the  density of states respectively from DFT 
calculation of pristine black phosphorus. (b) and (d) show the band structure and the density of states for phosphorene.}
\label{fig2}
\end{figure*}
\begin{figure*}
(a)
\includegraphics[angle=0,width=0.5\columnwidth]{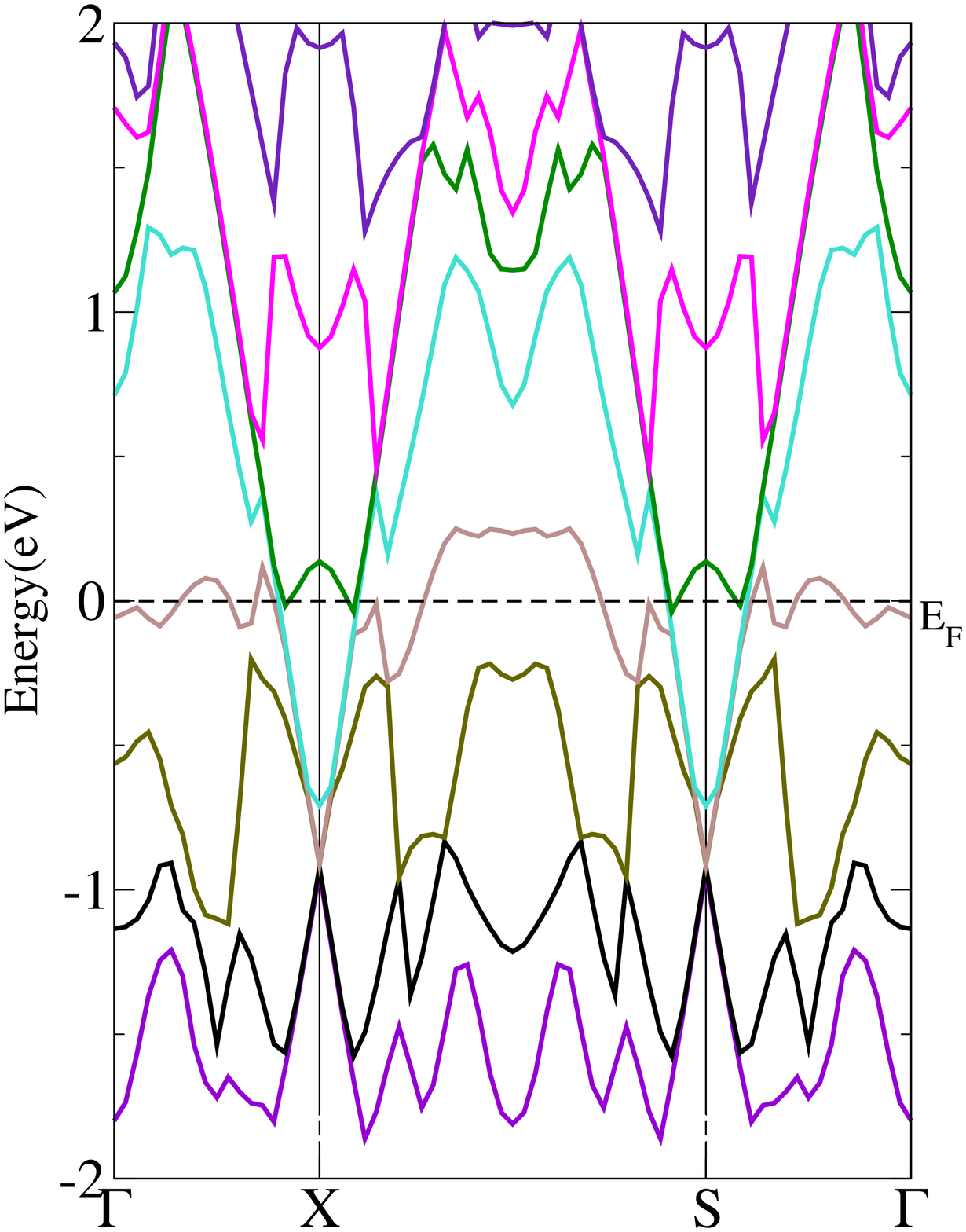}
(b)
\includegraphics[angle=0,width=0.5\columnwidth]{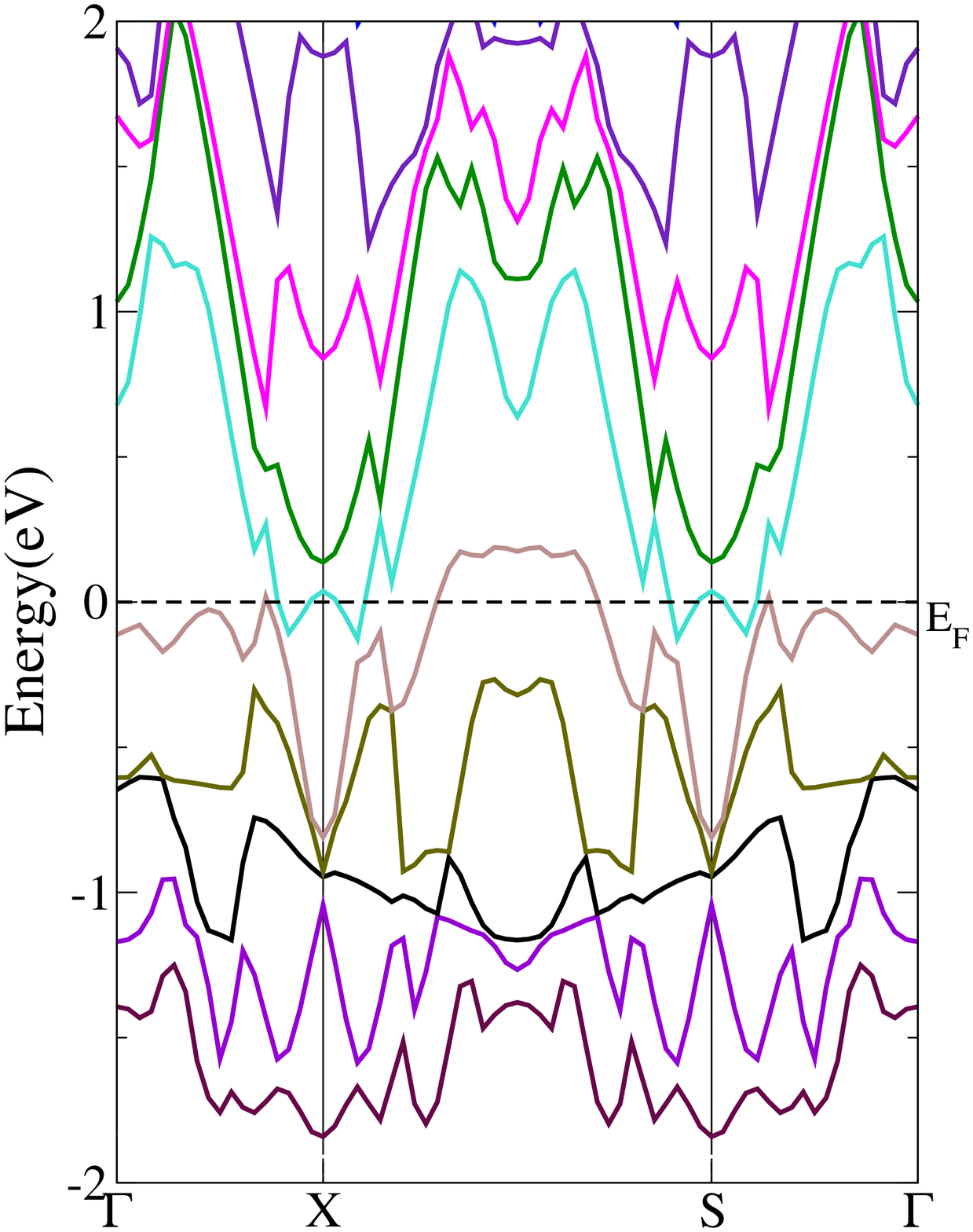}
(c)
\includegraphics[angle=0,width=0.5\columnwidth]{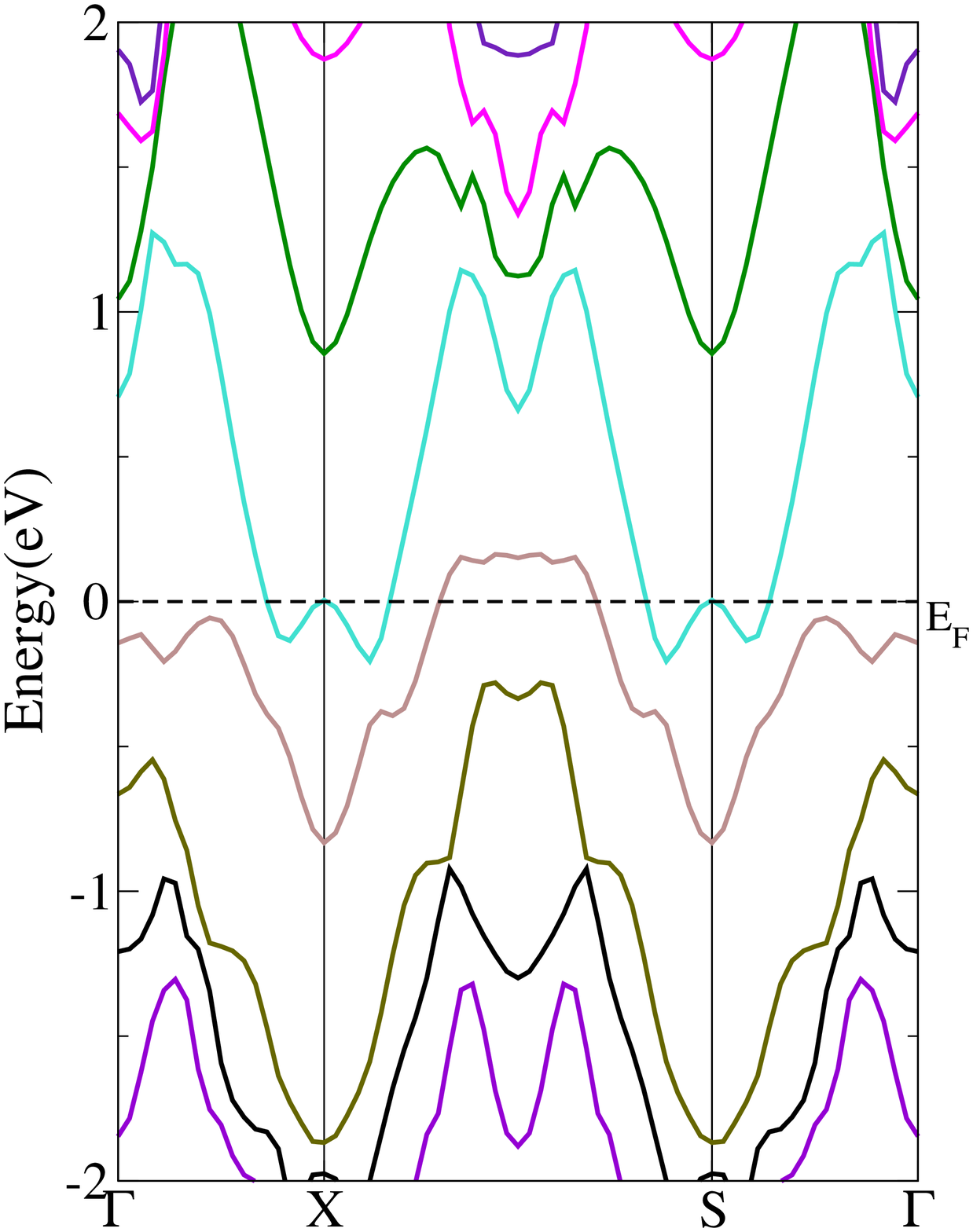}

(d)
\includegraphics[angle=0,width=0.5\columnwidth]{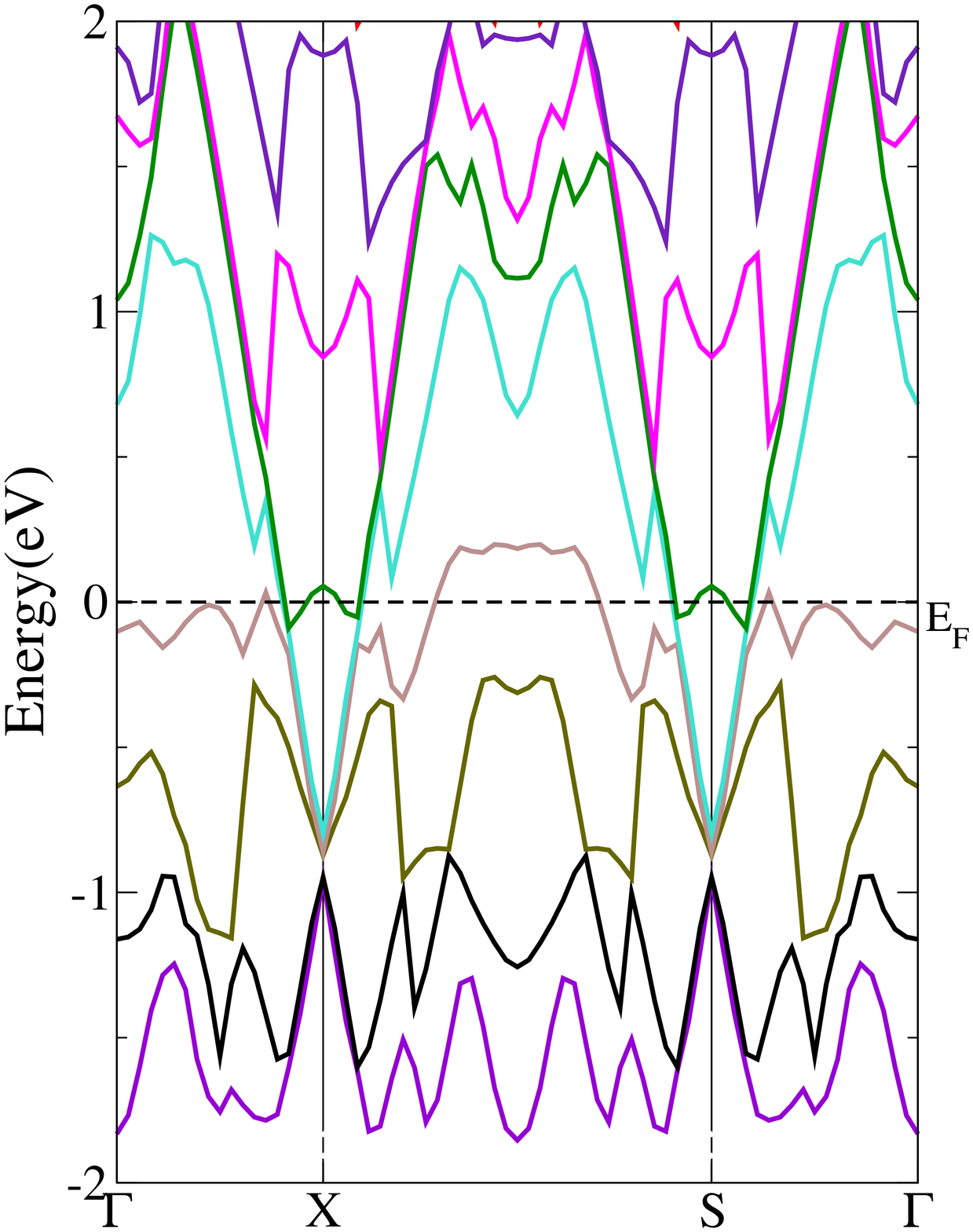}
(e)
\includegraphics[angle=0,width=0.5\columnwidth]{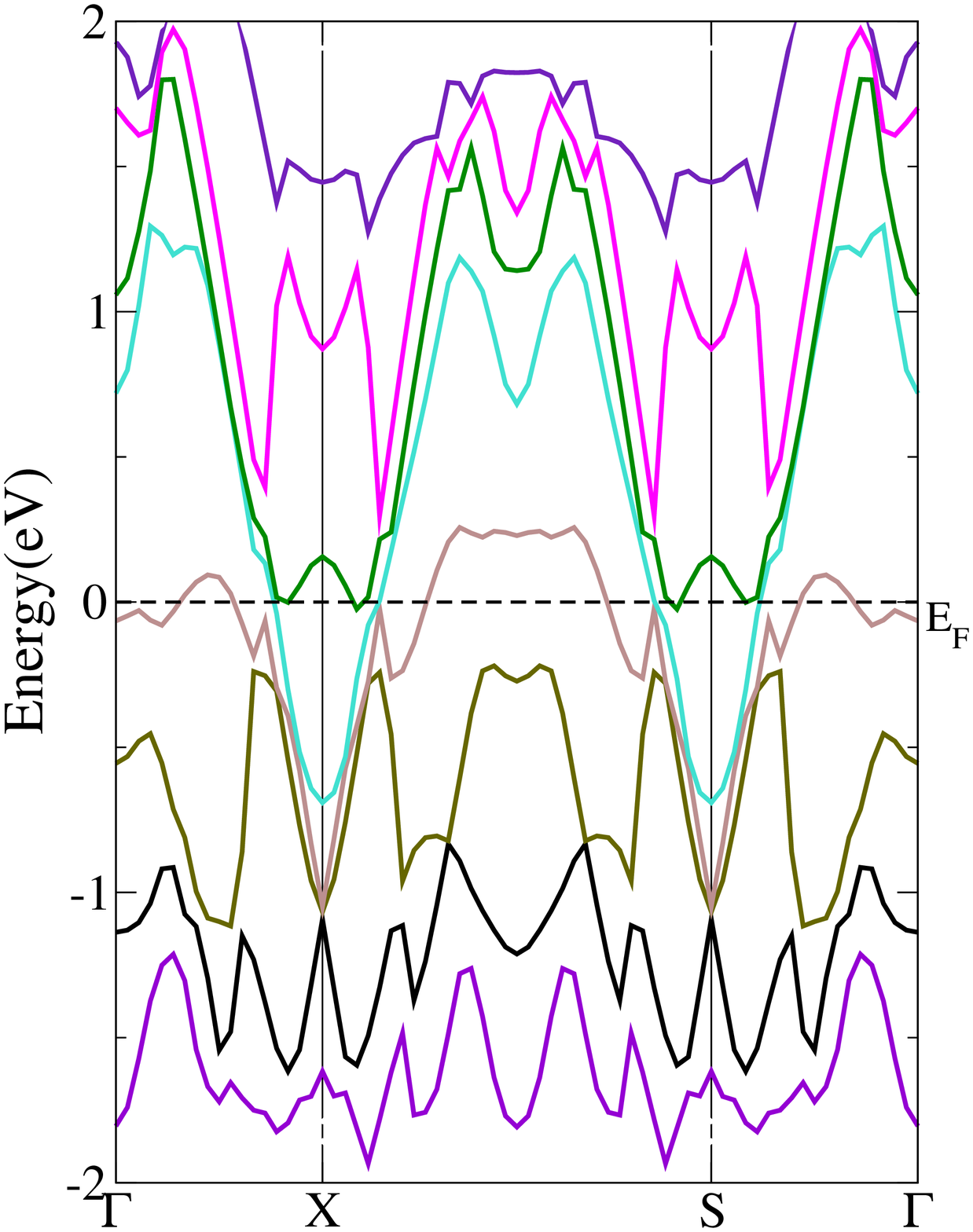}
(f)
\includegraphics[angle=0,width=0.5\columnwidth]{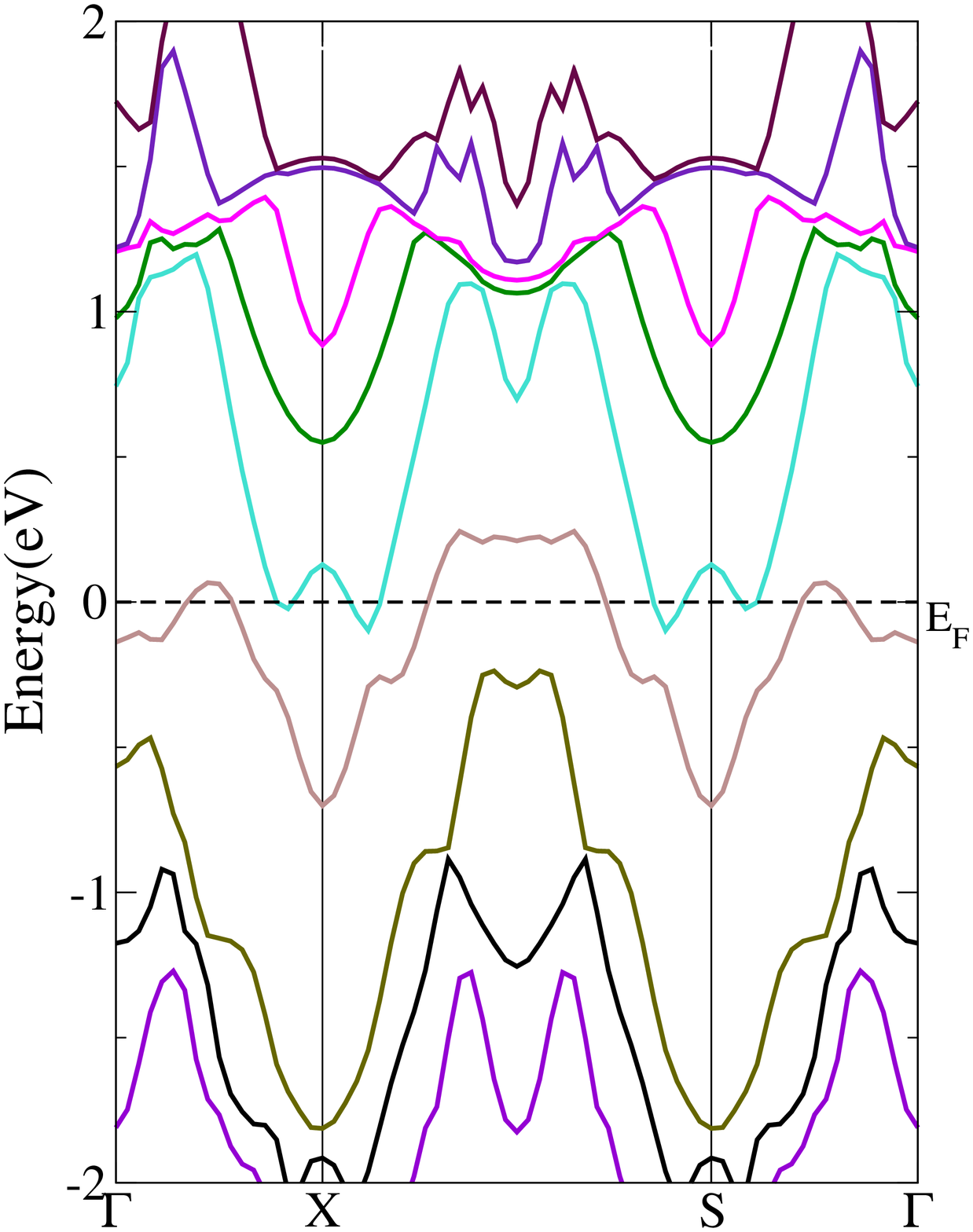}
\caption{(Color Online) Spin polarized DFT band structures for nitrogen 
intercalation. (a),(b) and (c) is for the up spin and others are for the down spin. 
The six plots are for intercalation concentration of 
(a),(d) 20 $\%$ (up and down spins respectively), (b),(e) 30 $\%$ (up and down spins respectively) and (c),(f) 40 $\%$ (up and down spins respectively).}
\label{fig3}
\end{figure*}

\begin{figure*}
(a)
\includegraphics[angle=0,width=0.5\columnwidth]{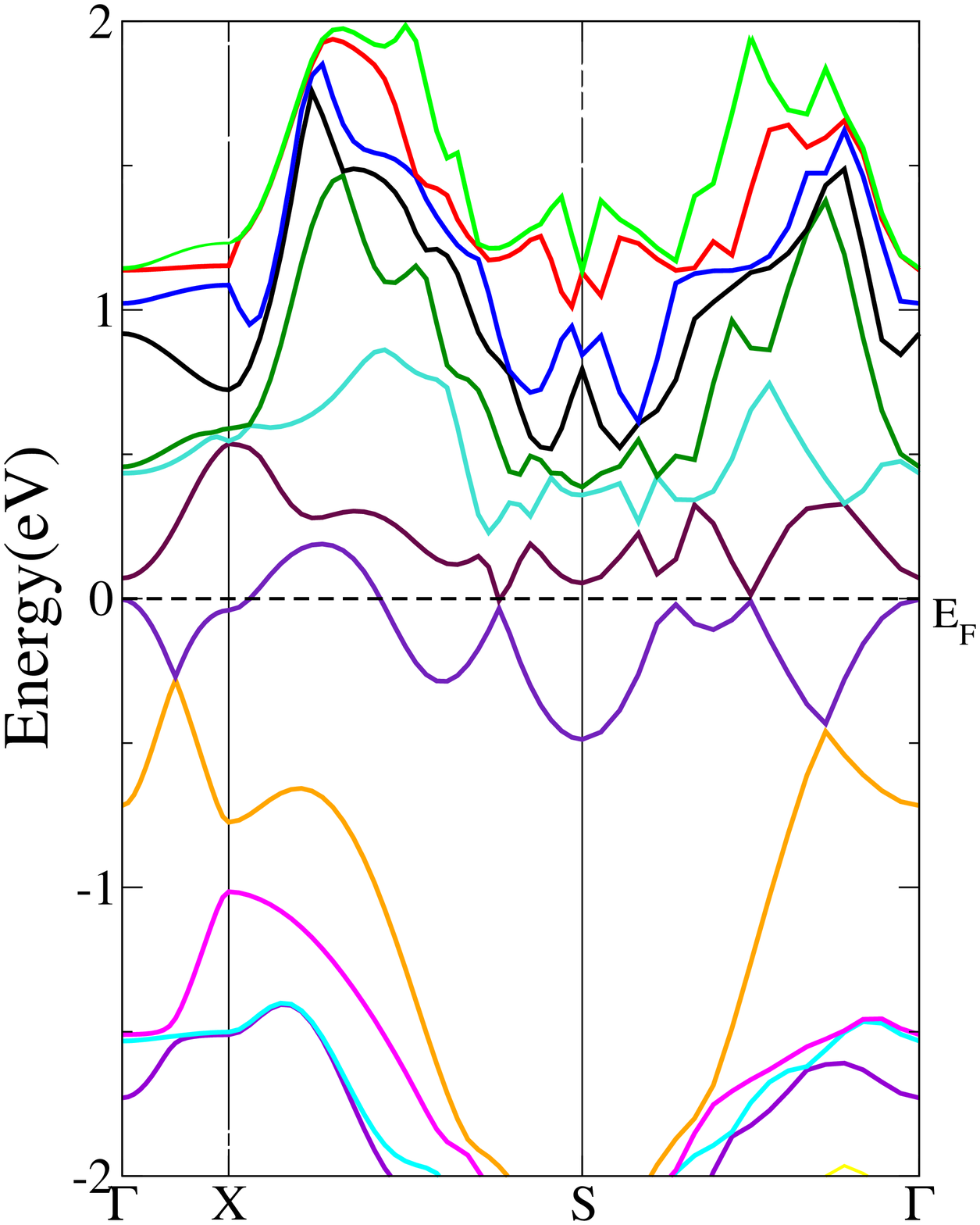}
(b)
\includegraphics[angle=0,width=0.5\columnwidth]{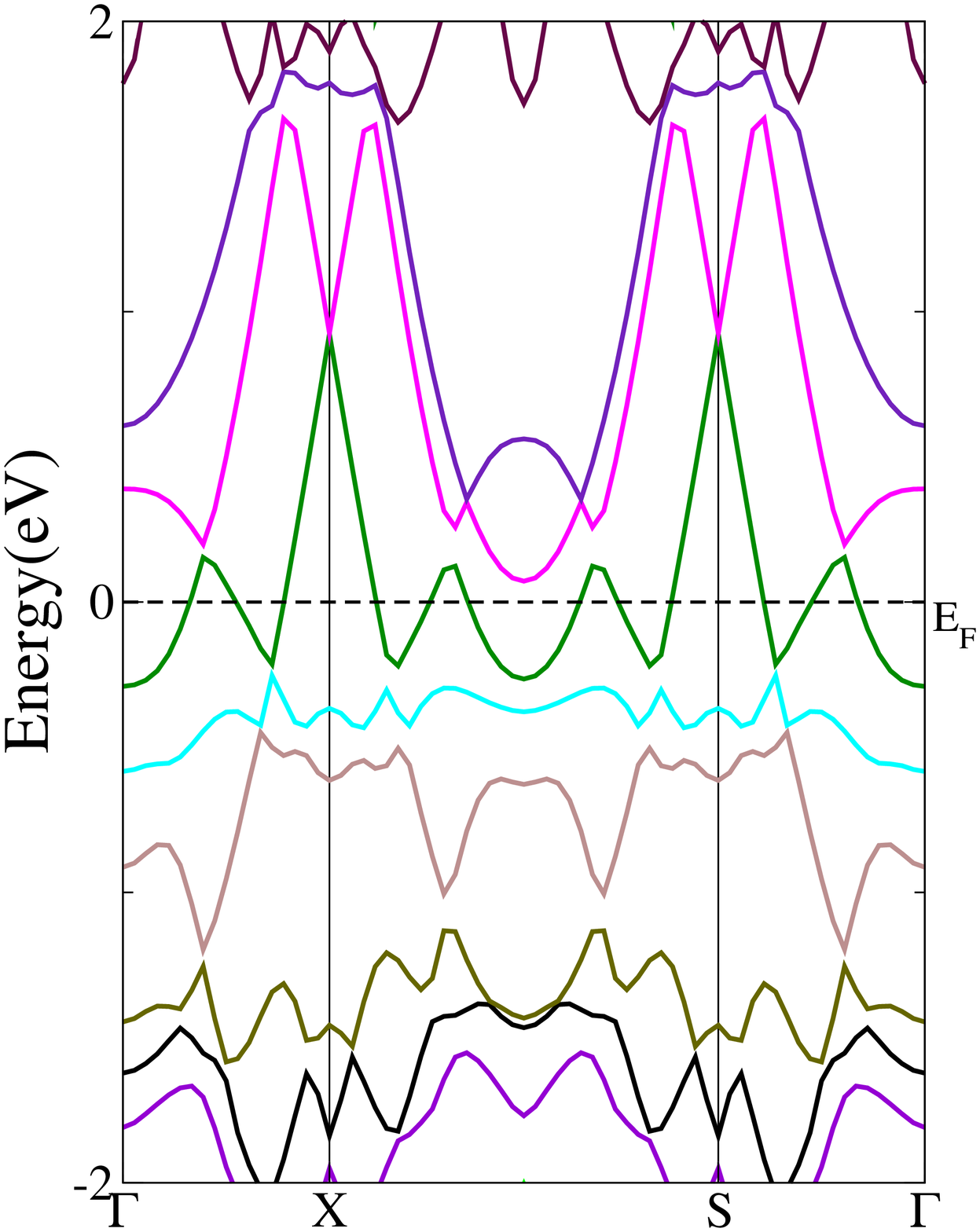}
(c)
\includegraphics[angle=0,width=0.5\columnwidth]{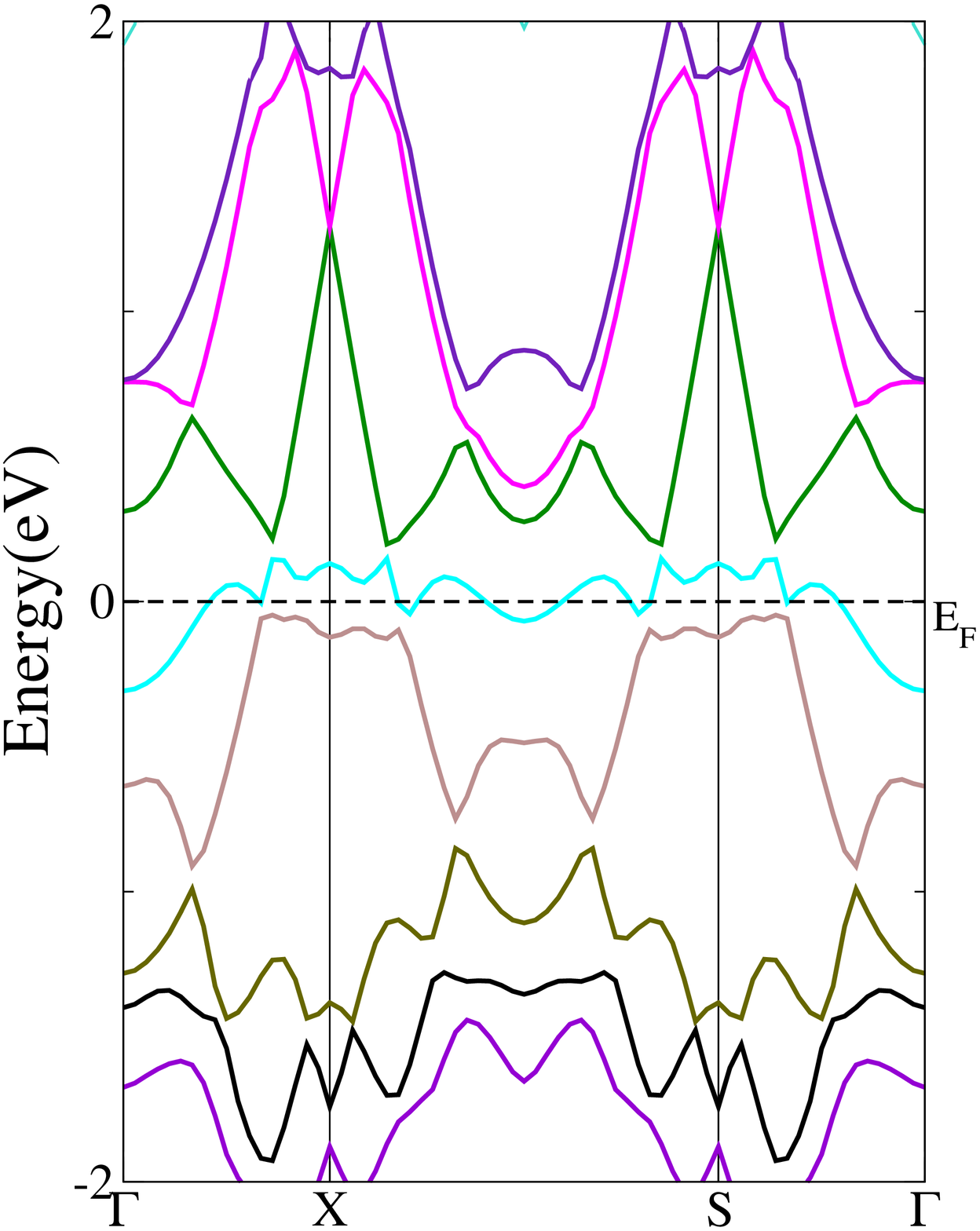}

(d)
\includegraphics[angle=0,width=0.5\columnwidth]{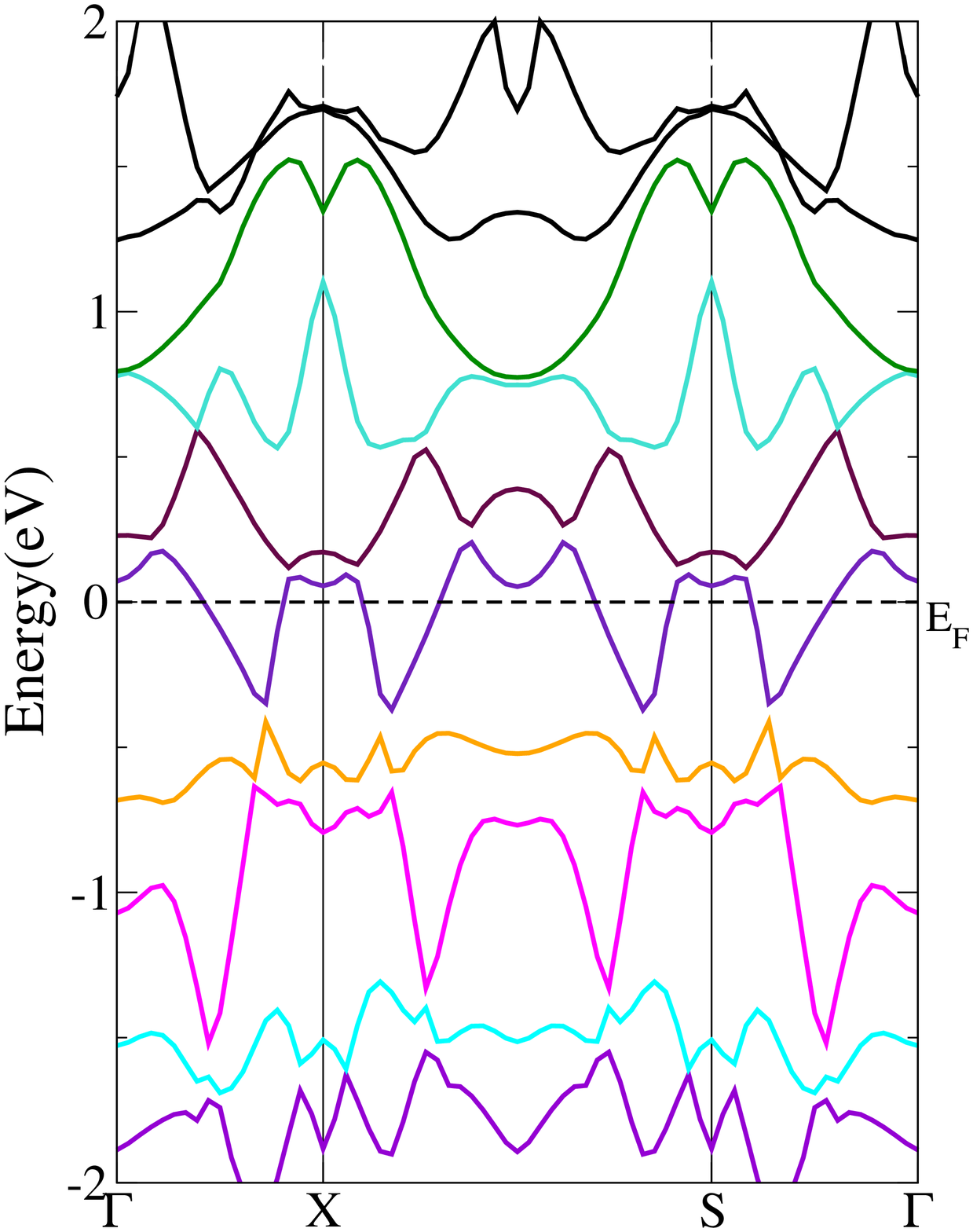}
(e)
\includegraphics[angle=0,width=0.5\columnwidth]{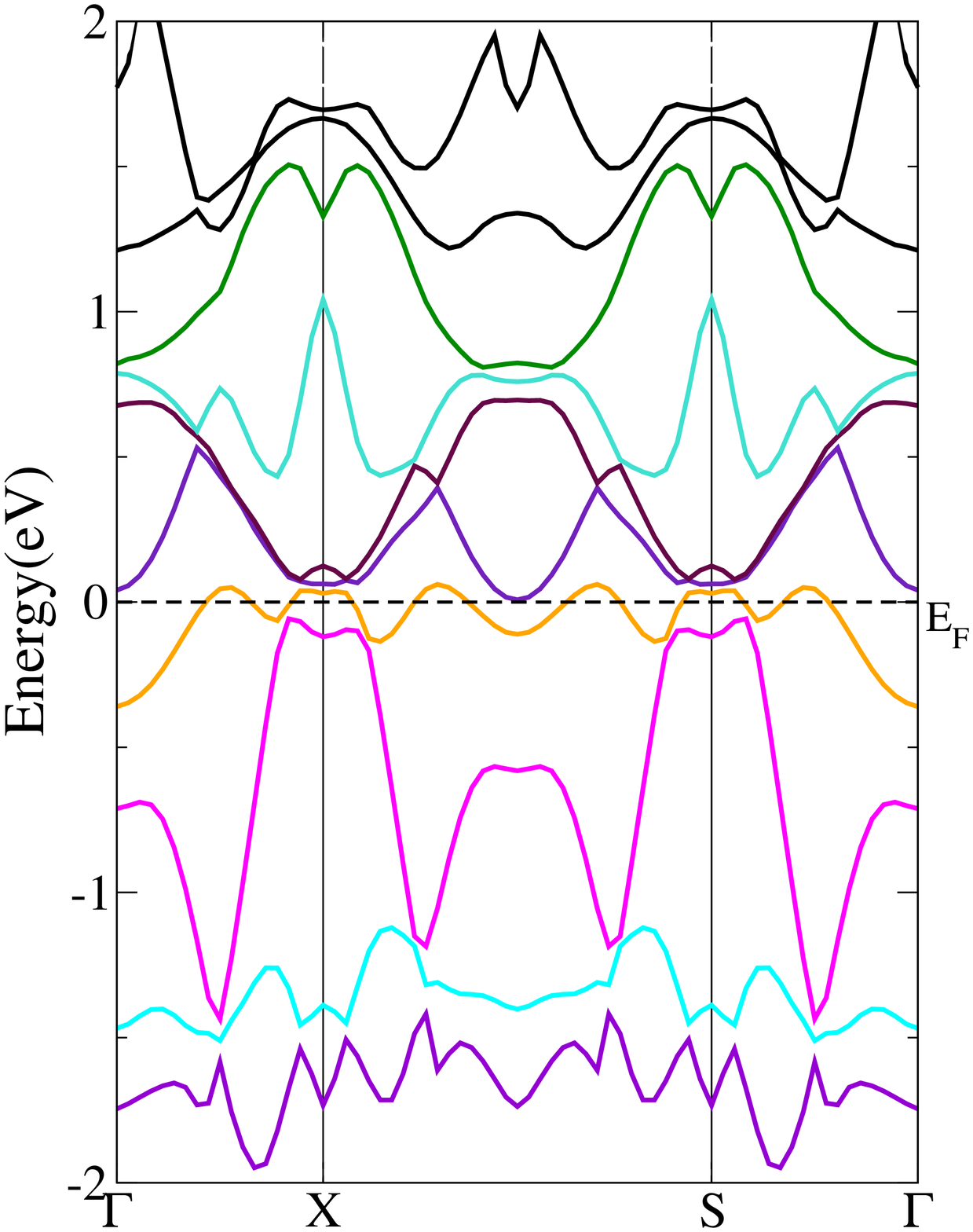}
\caption{(Color Online) Spin polarized DFT band structures for lithium 
intercalation. (a),(b) and (d) is for up spin and others are for down spin. 
The plots are for intercalation concentration of 
(a) 20 $\%$ (up spin), (b)-(c)30 $\%$ of lithium (up and down spins respectively) and (d)-(e)20 $\%$ of calcium (up and down spins respectively).}
\label{fig4}
\end{figure*}
The above findings inspire us to study the spintronic properties of
intercalated phosphorene and compare them to the earlier studies on phosphorene 
since doping induced magnetization is already established.
We further show that the intercalation induced changes for other intercalant 
atoms, such as, lithium and calcium. 
It may be noted that doped phosphorene have already been shown to be a promising candidate for spintronics\cite{yu}.
Here we investigate intercalation with nitrogen, lithium and calcium, and in particular, large concentration of
lithium and nitrogen yield effective spintronic materials. To the best of our knowledge studies on the intercalation induced 
spintronic property have not been reported in literature.

In this paper, we use density functional theory to investigate the electronic 
and magnetic properties of intercalated phosphorene.  Subsequently we show that though 
phosphorene has a band gap, the process of intercalation reduces the band gap and
it induces magnetic properties. In the following sections we first describe our 
methodology for calculating the band structures and the density of states. Hence we 
analyze and discuss the results for each of the intercalants and their contribution to
spintronic properties. A brief summary finally follows.

\section{Computational details}
We studied intercalation of phosphorene with nitrogen, lithium and calcium as a function of 
the intercalation densities to evaluate their
density and the atomic number dependencies on the electronic and 
magnetic properties. We have compared our results on the
intercalated phosphorene with that of the pristine one. 
First-principle methods based on density functional theory 
(DFT)\cite{dft} are employed for this purpose. The computations are performed using WIEN2K full-potential 
linearized augmented plane wave(FP-LAPW)\cite{wien2k} ab initio package 
within the DFT formalism to get the electronic band structures and corresponding total density of states 
(DOS). A generalized gradient approximation Perdew-Burke-Ernzerhof (GGA-PBE) 
exchange correlation potential is used here. In order to find the magnetic 
properties, spin-polarized calculations are implemented on the intercalated 
phosphorene. For the pristine phosphorene we used a unit cell of 4 P atoms and 
for intercalation we used a $2\times2$ supercell of this material. Intercalants 
are placed in the middle of the two layers to check for the convergence and the energy minimized 
structure. The parent crystal structure and the intercalated structure are shown in 
Fig.1. The convergence limit is set to 0.0001 eV for the kinetic energy 
cutoff and the $k$-point mesh. The muffin-tin radii, $R_{MT}$ is chosen as 2.09 for 
phosphorus, 1.92 for nitrogen, 1.56 for lithium and 2.45 for calcium atoms. 
With the choice of $Rk_{max}$ ($Rk_{max}$ stands for the product of the smallest
 atomic sphere radius, $R_{MT}$ times the largest $k$-vector $k_{max}$) as 7.0 and 
a $10\time10\times10$ $k$-mesh, the total energy is calculated while relaxing all 
the atoms so that the maximum force is smaller than 0.01 $eV$/$\AA$. Then the band 
structure, the density of states and the total magnetization are calculated from the
converged spin polarized self-consistent field (scf) calculations.

\section{Results and Discussion}
The band structure of parent black phosphorus shows a band gap and 
nearly a Dirac like feature along the Z-$\Gamma$ direction of the brillouin zone. 
Here we plot the band structure along the high symmetry $k$-points in the first 
Brillouin zone of the reciprocal space along the $\Gamma$-X-L-Z-$\Gamma$ 
direction. The results are shown in Fig.2a. The corresponding DFT density of states is presented in Fig.2c, 
which shows the presence of a gap at the Fermi energy. In Fig.2b and 2d we 
have shown the band structure and the density of states of monolayer phosphorus which shows a gap of 0.91 eV which matches with earlier
 GGA-PBE calculation\cite{1404.5171}. The band structures of the pristine black phosphorus also agrees with existing results\cite{1404.5171}. 
We have performed spin-polarized calculations for the intercalated materials 
and the corresponding band structures for the up and the down spins are obtained. The results appear in 
fig.3a-3f for the nitrogen intercalation, while those for the lithium and the calcium intercalation are shown in 
fig.4a-4e.

Let us focus on Fig.3 which shows the band 
structure of nitrogen intercalated phosphorene. The results for the up and the down spin
densities of states appear in three columns for three different nitrogen concentrations, namely, 
20$\%$, 30$\%$ and 40$\%$.  When compared with the corresponding results for pristine
phosphorene, it may be concluded that the intercalation 
changes the band structure considerably both quantitatively and qualitatively. The intercalation 
induces lowering of the overall gap, while gaps between the
up-spin and the down-spin channels evolve. 
\begin{figure*}
(a)
\includegraphics[angle=270,width=0.6\columnwidth]{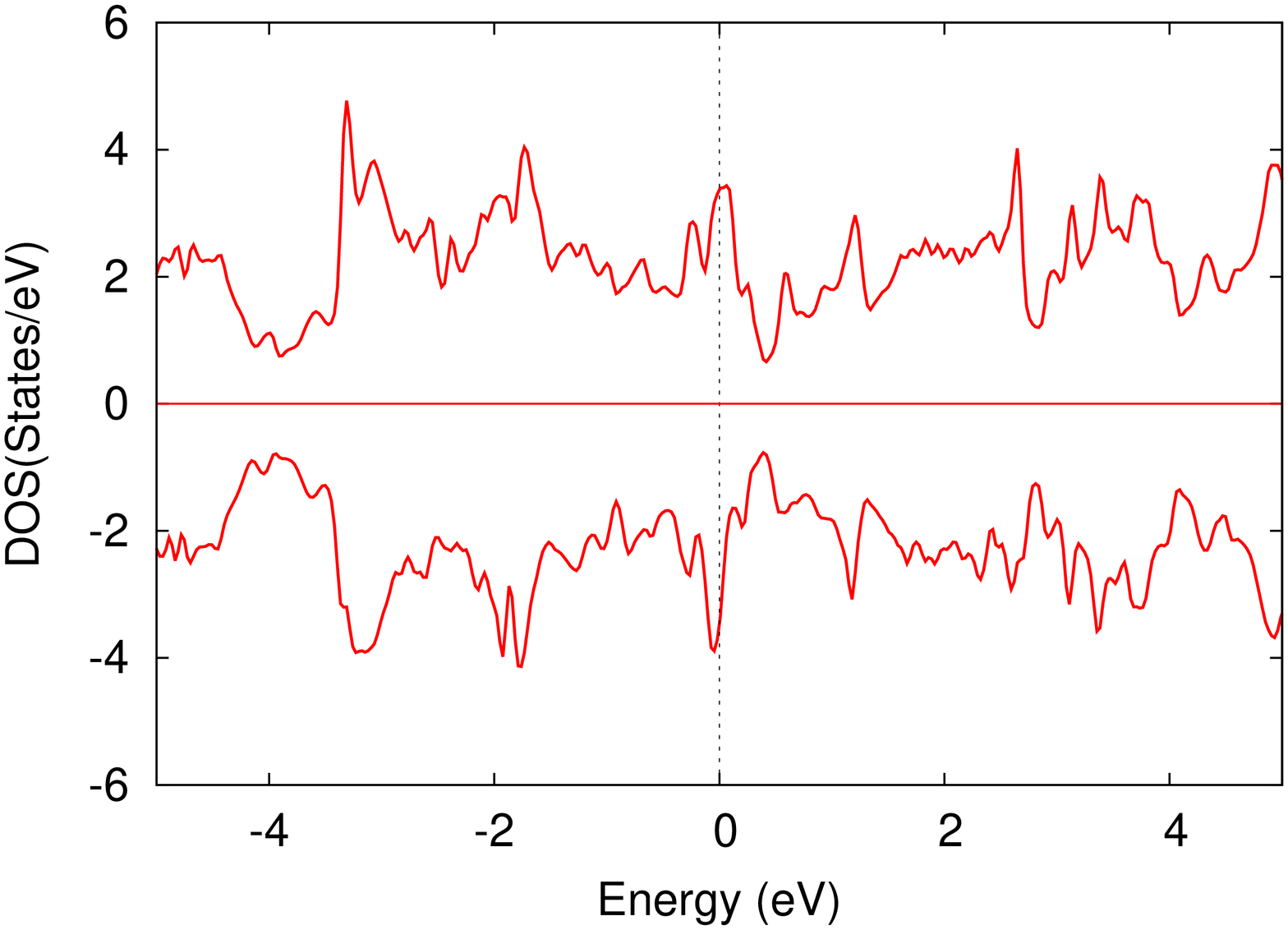}
(b)
\includegraphics[angle=270,width=0.6\columnwidth]{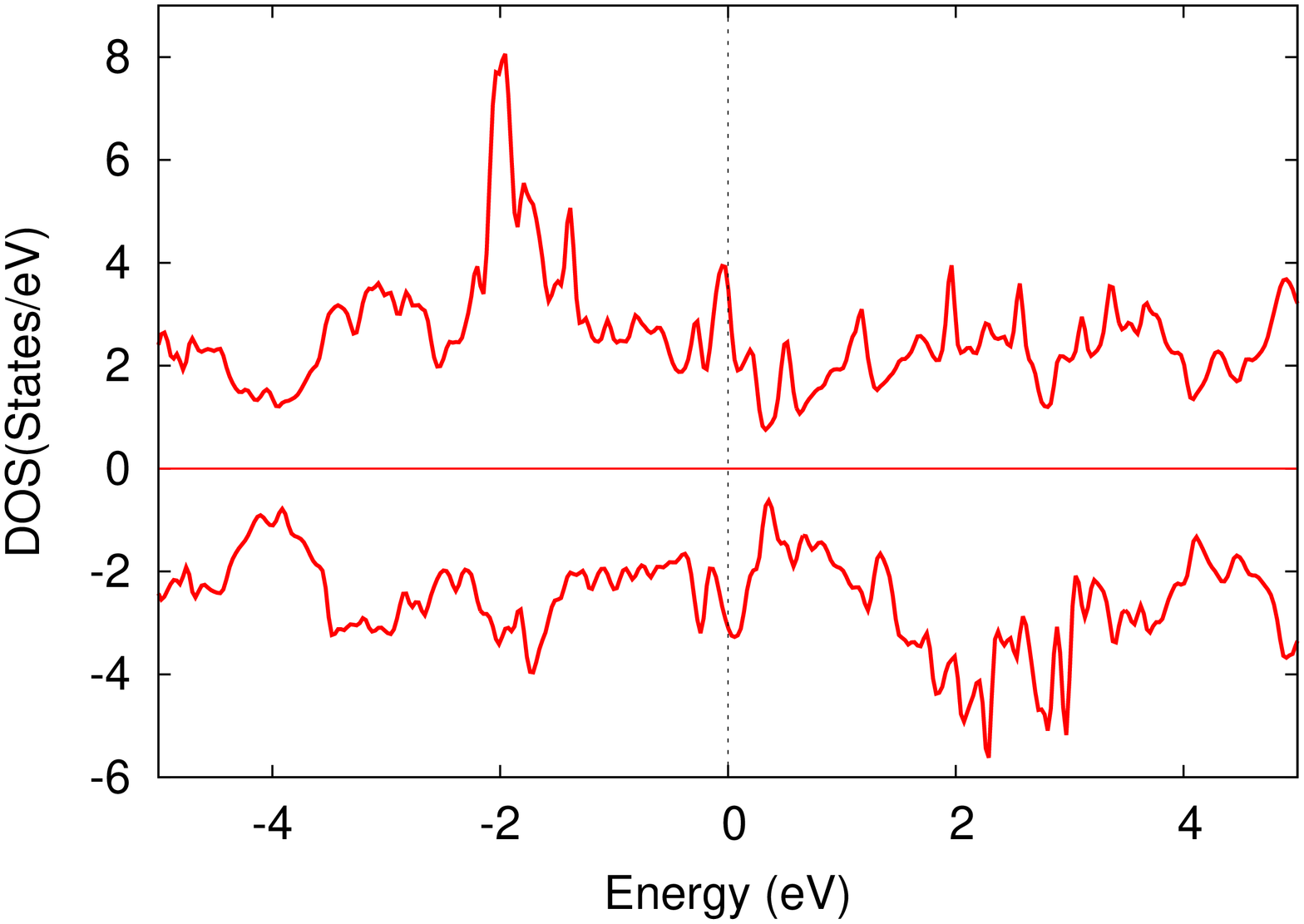}
(c)
\includegraphics[angle=270,width=0.6\columnwidth]{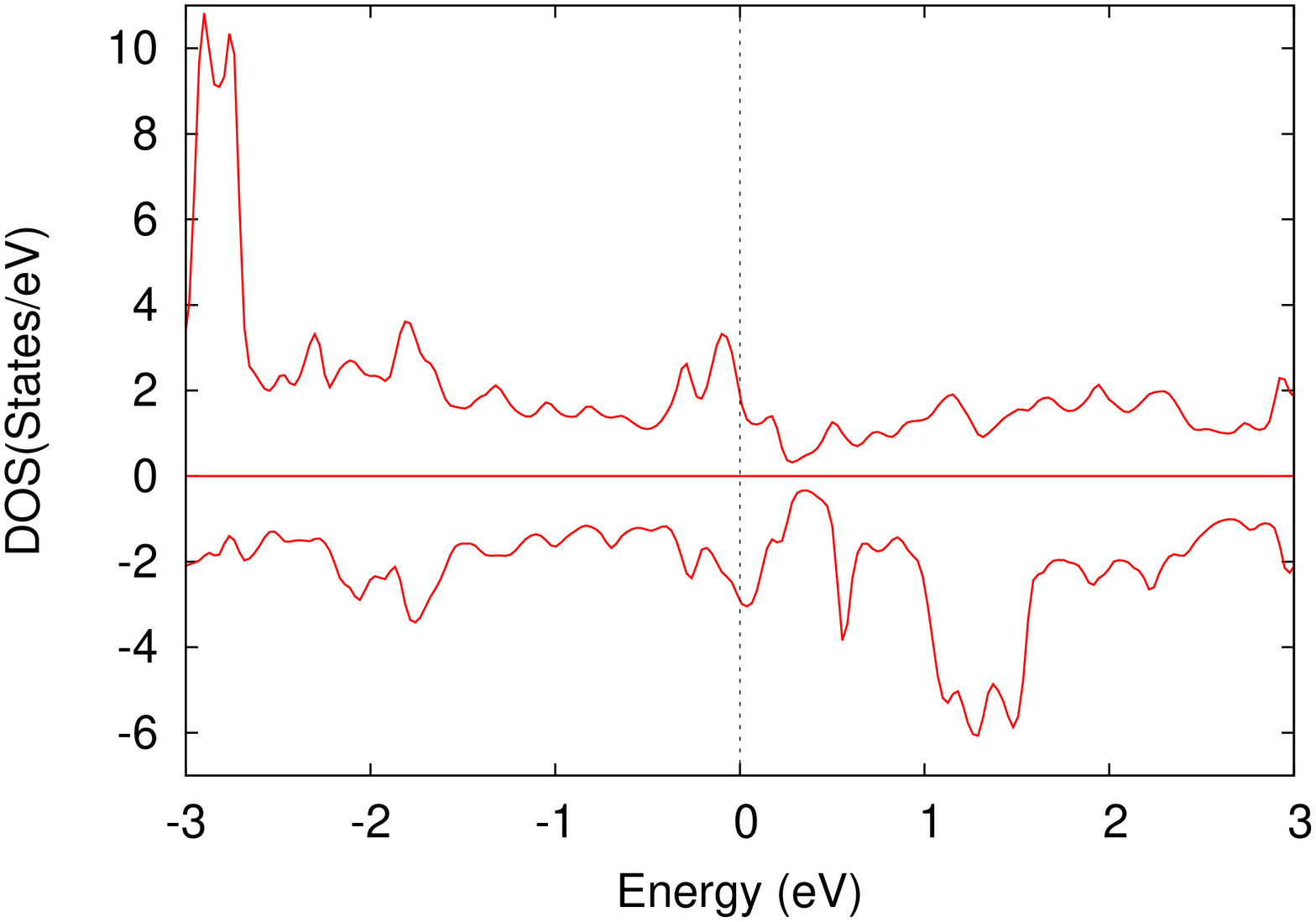}
\caption{(Color Online) DFT density of states with nitrogen intercalation. Upper panel is for up-spin and lower panel is for down spin, down spin density of 
states is multiplied by -1. The three plots are for intercalation concentration of (a) 20 $\%$, 
(b) 30 $\%$ and (c) 40 $\%$.}
\label{fig5}
\end{figure*}
\begin{figure*}
(a)
\includegraphics[angle=270,width=0.6\columnwidth]{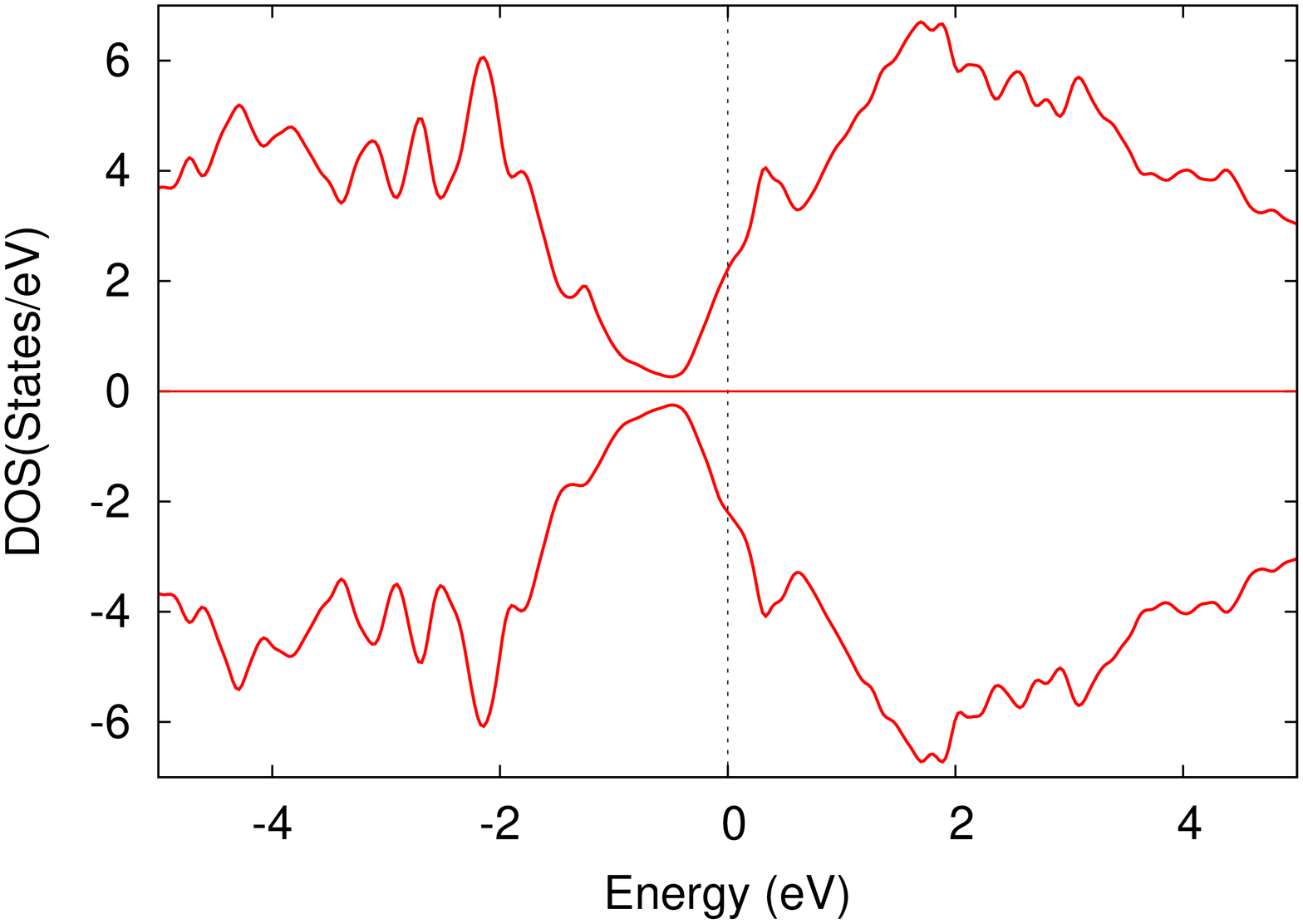}
(b)
\includegraphics[angle=270,width=0.6\columnwidth]{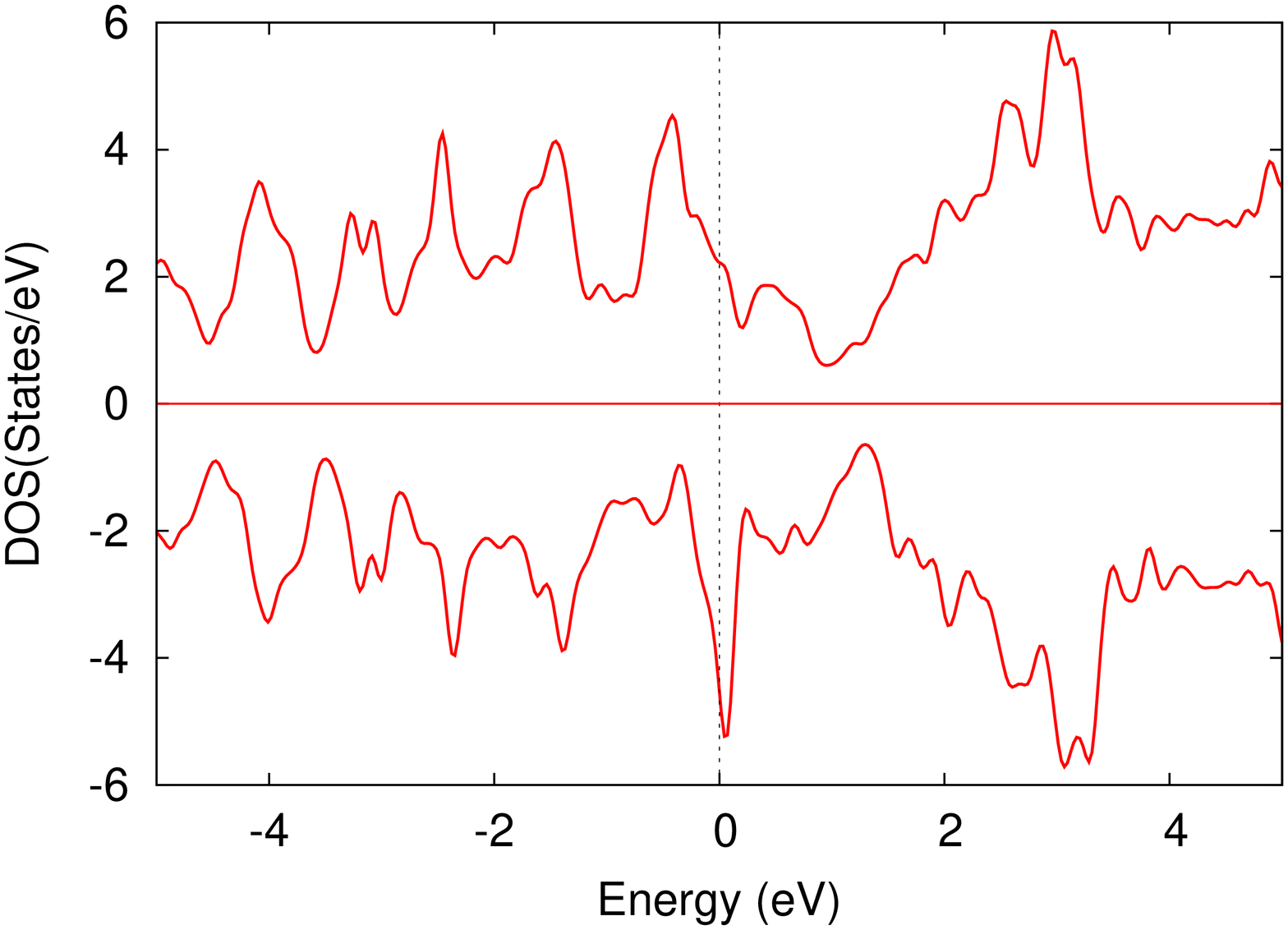}
(c)
\includegraphics[angle=270,width=0.6\columnwidth]{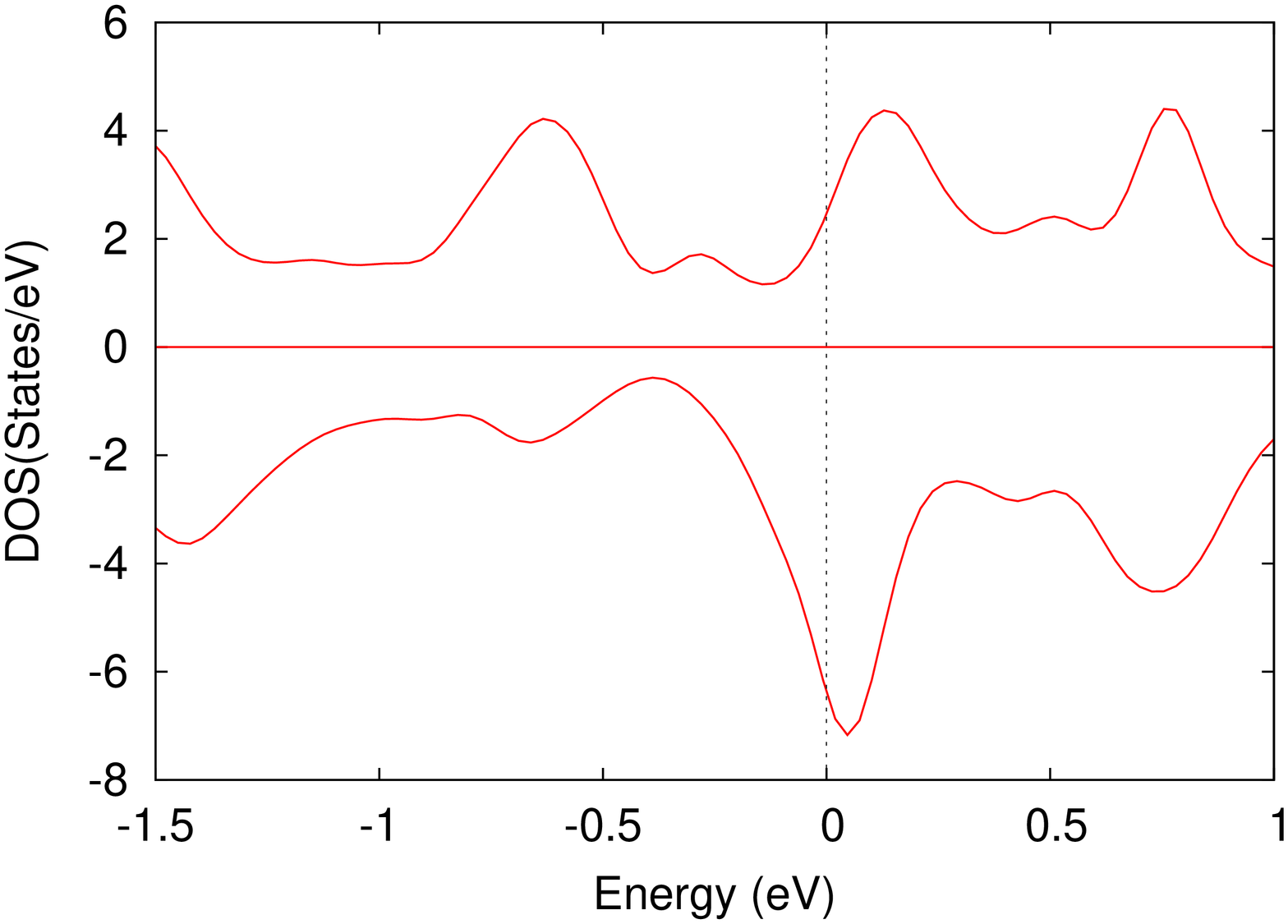}
\caption{(Color Online) DFT density of states corresponding to lithium and calcium 
intercalation. Upper panel is for up-spin and lower panel is for down spin. The down spin density of 
states is multiplied by -1. The three plots are for intercalation concentration of (a) 20 $\%$ 
and (b) 30 $\%$ of lithium and (c) 20 $\%$ of calcium.}
\label{fig6}
\end{figure*}
\begin{figure}
(a)
\includegraphics[angle=270,width=0.8\columnwidth]{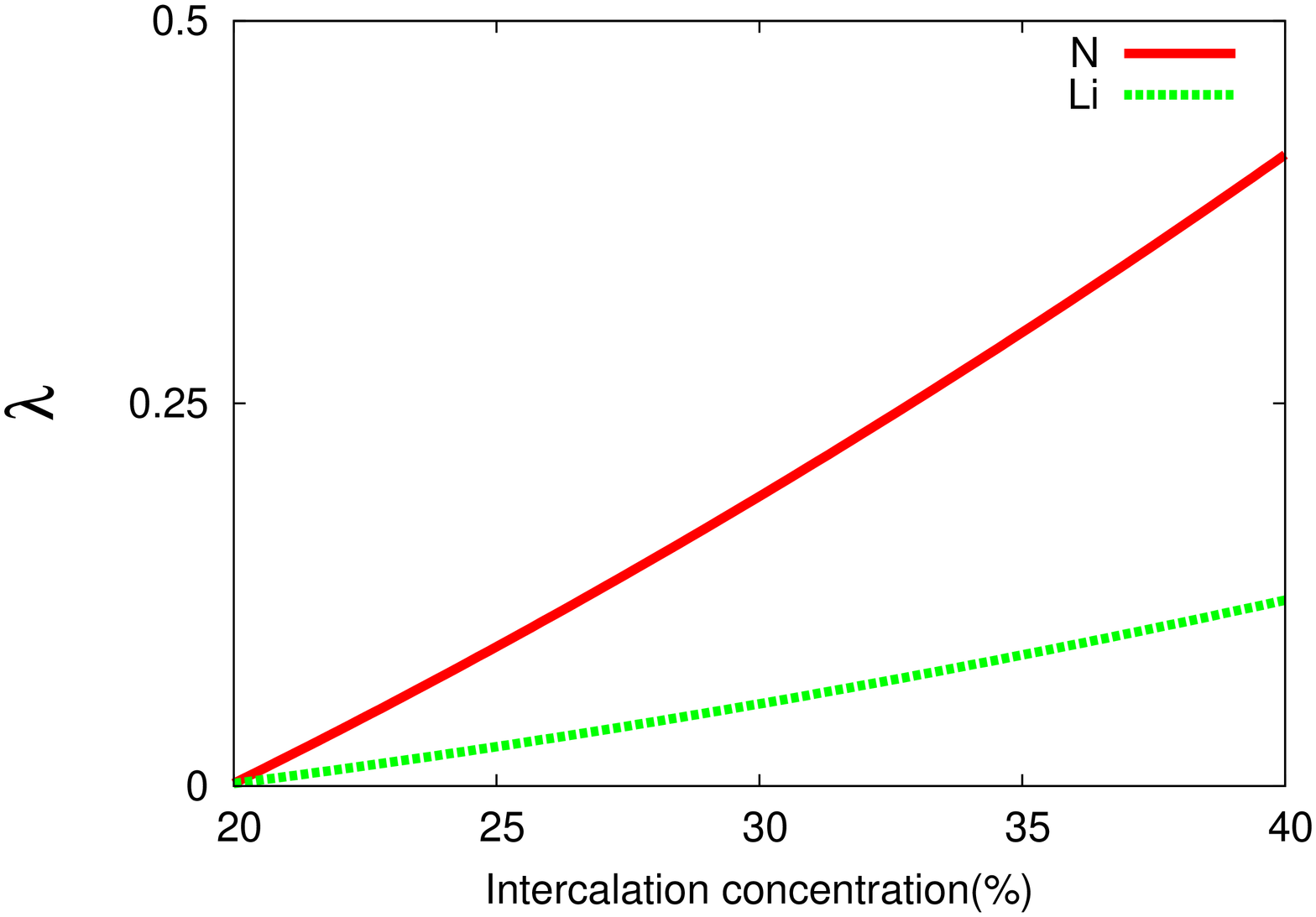}

(b)
\includegraphics[angle=270,width=0.8\columnwidth]{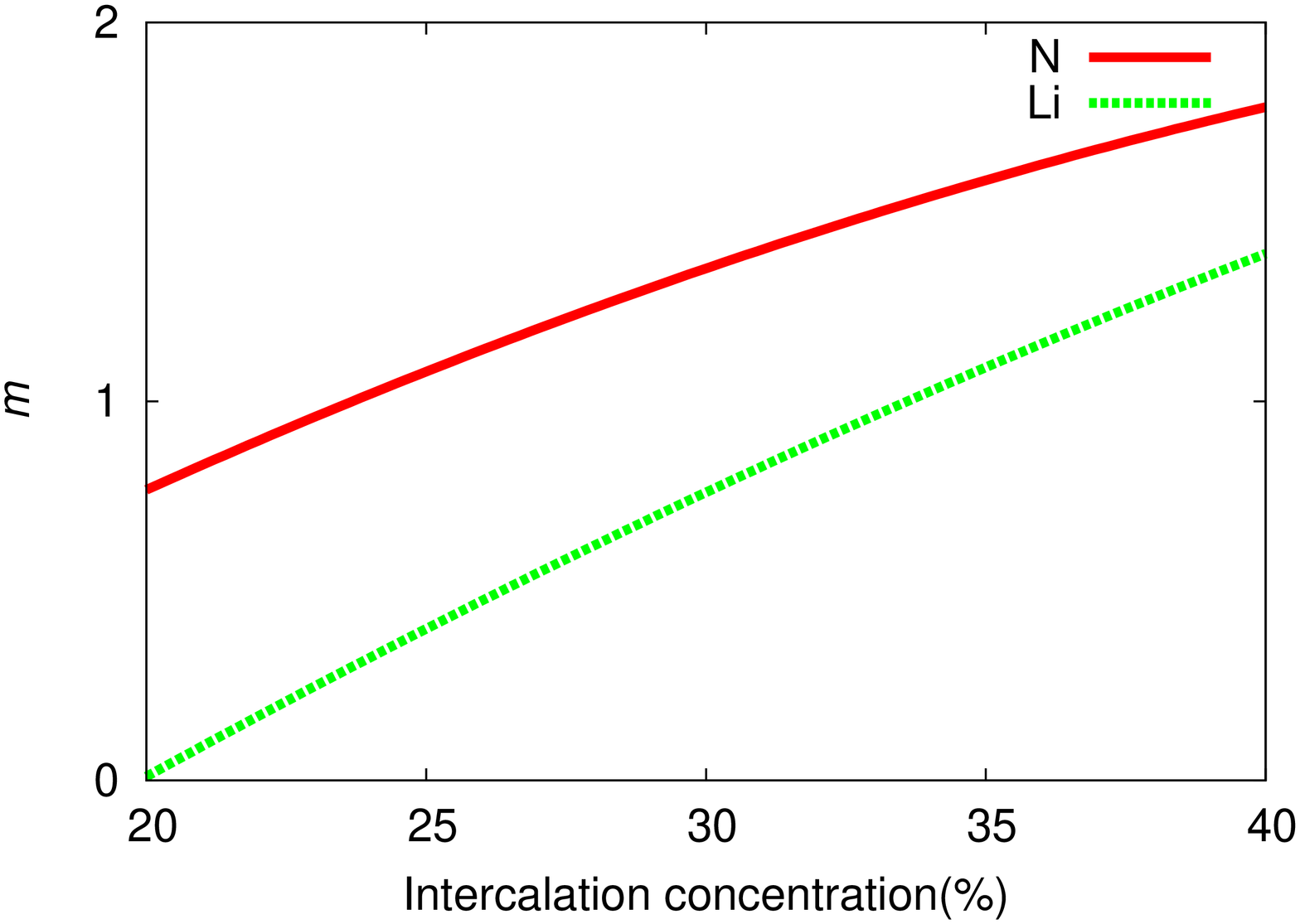}
\caption{(Color Online) (a) Spintronic order parameter ($\lambda$) plot with intercalation 
concentration. It shows spintronic order parameter increases with intercalation 
density. (b) Plot of magnetization, $m$=(n$_{\uparrow}$-n$_{\downarrow}$)$\mu_{B}$\cite{data} variation with the intercalation concentration. Here we have taken $\mu_{B} =1$.}
\label{fig7}
\end{figure}

%

To further check for the concentration dependence of different elements on its spintronic properties 
we have used lithium and calcium as intercalants. In both the cases, the band structures
 and the densities of states are calculated to enumerate their features. For lithium,  
we have used 20$\%$ and 30$\%$ as the intercalation concentration, and for calcium we
 have used only 20$\%$ as higher concentrations are found to be energetically unfavourable.
We have observed that for 20$\%$ lithium intercalation, the spin polarized band structure 
shows no gap between the up and down spins, while upon increasing the concentration, a 
gap emerges. Thus we have presented the band structure for one type of spin in Fig.4a and 4b 
and in Fig.4c, we have presented the band structures for both the up and the 
down spins respectively, at
higher values, that is, for 30$\%$ lithium intercalation.
It is evident that there is a considerable change between the up and down spin densities.
So intercalating with lithium needs larger concentration in order to obtain pronounced  
spintronic properties. Interestingly, intercalation has also changed the band structure 
considerably  compared to the DFT band structure, which implies that the impurity bands have 
more pronounced effects near the Fermi level, and the band gap diminishes with intercalation.

In contrast to lithium and nitrogen, 20$\%$ of calcium intercalation in phosphorene 
makes a large change in the band structure  and the difference between the up and the
down spins is also high for this material. In the band structure it can be 
observed that, though there is no gap at the Fermi level, the bands near the Fermi 
level do not overlap with each other for the up spin which carries a signature that 
tuning the Fermi level with external voltage can make the system a gapped one. 
However higher concentration of calcium intercalation is energetically 
forbidden due to the size mismatch of the calcium atoms with that of the pristine 
phosphorus, which constitutes a major drawback of calcium intercalation at large densities.  

Spin polarization induces a feature around the
$\Gamma$ point. The Dirac-like property along the Z-$\Gamma$ direction vanishes with 
intercalation. The difference in the band structures for the up and the down spins
indicates that the intercalated phosphorene is a promising candidate for spintronics applications. To get a better understanding of the difference between
the up and the down spin density of states (DOS) we present here the spin-polarized DOS  for the
intercalated phosphorene in Fig.5a-5c for nitrogen intercalation with different values of
concentration, and Fig.6a-6c for lithium and calcium intercalation respectively. As 
indicated from the band structure, the DOS 
also shows gapless feature in the up and the down spin channels, while there is a
sharp difference between the DOS for the up and the down spins for three different nitrogen 
densities. Whereas, for low lithium intercalation, the up and the down spins 
show similar structures as predicted also from the band structure plots. However with increased concentration, there is a difference 
between the up and the down spin density of states. Similarly, the calcium intercalation also shows 
a large  difference in the densities of states for the two spins.  In these spin polarized systems 
we found that the source of polarization is from the intercalation and the major contribution
to the magnetic moment is due to the intercalating material. These results agree 
with the calculated band structures and the data for the spin densities. 

We wish to reiterate that cases of large intercalation densities are permissible
 owing to the stability of the structures at high values of the intercalant 
densities, except for calcium. We ensure this via plotting the energy as a 
function of the number of iterations used in our scf method. Clearly a stable 
minimum in each of the plots shown in Fig.1c is a testimony of that. In addition to providing the structural stability of the intercalated compounds, it allows devising more efficient spintronic materials for reasons we elaborate below.

\begin{table}
\begin{tabular}{|c|c|c|c|c|}
\hline
Intercalant  & Atomic No  & $\lambda$\\
\hline
lithium & 3 & 0.0004588\\
\hline
nitrogen & 7 & 0.01715  \\
\hline
calcium & 20 & 0.12 \\
\hline
\end{tabular}
\caption{Values of spintronic order parameter $\lambda$ for intercalation in parent 
phosphorene with lithium, nitrogen and calcium with an intercalation concentration of 20$\%$. }
\label{table1}
\end{table}

The performance of a spintronic material is determined by the 
degree of spin polarization of the carriers of a system. To firmly establish 
the degree of polarization of the charge carriers in these materials, 
we have computed the 
spintronic order parameter ($\lambda$) which is defined by, 
\begin{equation}
\lambda = \frac{n_{\uparrow} - n_{\downarrow}}{n_{\uparrow} + n_{\downarrow}}
\end{equation}
where $n_{\uparrow}$ and $n_{\downarrow}$ are the occupation densities of the up and the down spins.
Larger values of $\lambda$ imply larger spin polarization and hence a better candidate 
for a spintronic material.
Here we show that $\lambda$ increases with increasing the intercalation density 
for nitrogen and lithium (fig.7a). Similar features emerge for 
$\lambda$ 
parameter as a function of the atomic number of the intercalants,  
thereby signaling emergence of improved spintronic properties upon intercalation
with elements having larger atomic number. The results are shown in table 1 which indicates that
as lithium is replaced by calcium, there is an enhancement of the value of $\lambda$ by three
orders of magnitude.  While we find this result remarkable and the study 
probably warrants 
intercalation with elements having larger atomic number, however more work is required to 
validate this claim as some of the larger elements may not yield an energetically favourable 
scenario.  Even pristine phosphorene is believed to be a promising 
candidate for spintronic applications, however intercalation certainly enhances the prospects of its applicability\cite{nitro}. 

In comparison with other spintronic materials, the intercalated phosphorene 
does not possess any gap in its band structure, however there is difference 
between the densities of the up and the down spin states, which, in effect induces 
magnetic properties in the system. The results are shown in Fig.7b where the 
magnetization, $m = (n_{\uparrow} - n_{\downarrow})\mu_B$ is plotted as a function of intercalataion
densities for nitrogen and lithium. The behaviour shown clearly suggests of a sub-linear increase in magnetization values with densities in both the cases.

Inspired by the spintronic properties of zero-gap 
materials\cite{X-L-wang}, the concept of a gapless spintronic semiconductor material has been 
put under scrutiny. Based on the zero-gap state there are unique and superior 
properties of these intercalated phosphorene in terms of: (i) zero energy is required to 
excite electrons from the valence band to the conduction band, (ii) the charge carriers,
both holes and electrons are fully spin-polarized, 
(iii) fully spin-polarized electrons or holes can be separated easily in spin 
Hall effect, (iv) manipulation in the Fermi level can be done independently for each 
spin channel easily using a gate voltage control. So the physical properties can 
be changed by external means, such as, pressure, electric field, electromagnetic radiation, 
impurity doping, magnetic field etc.

In conclusion, we have observed that the phosphorene monolayer, which is an
anisotropic direct gap semiconductor with high carrier mobility, exhibits strong 
magnetic properties which scales with the 
intercalation density and the choice of the intercalant. These properties make them 
exceptional materials for zero band gap spintronic devices. The density dependence of the magnetic properties 
indicates flexible control on the magnetization of phosphorene. 
This opens up new possibilities of engineering room-temperature 
magnetic and spintronic devices.

\section{Acknowledgement}
SK acknowledges DST women scientist grant SR/WOS-A/PM-80/2016(G) for financial aid. SB thanks SERB, India for financial support under grant
no. EMR/2015/001039.


\begin{thebibliography}{60}
\bibitem{felser}C Felser {\it et al.}, {\it Angewandte Chemie International Edition} {\bf 46}, 668 (2007).
\bibitem{baibich}M. N. Baibich {\it et al.}, {\it Phys. Rev. Lett.} {\bf 61}, 2472 (1988).
\bibitem{sun}M. Sun {\it et al.}, {\it Appl. Surf. Sci.} {\bf 356}, 110 (2015).
\bibitem{JAP}J. S. Blakemore, {\it J. Appl. Phys.} {\bf 53}, R123 (1982).
\bibitem{natmet}P. Sharma {\it et al.}, {\it Nat. Mater.} {\bf 2}, 673 (2003).
\bibitem{ding}Y. Ding and Y. Wang, {\it J. Phys. Chem. C} {\bf 119}, 10610 (2015).
\bibitem{mak}K. F. Mak {\it et al.}, {\it Phys. Rev. Lett.} {\bf 105}, 136805 (2010).
\bibitem{tran}V. Tran {\it et al.}, {\it Phys. Rev. B} {\bf 89}, 235319 (2014).
\bibitem{choudhuri}I Choudhuri {\it et al}. {\it Chem. Mater.} {\bf 31}, 8260 (2019).
\bibitem{binaschprb}G. Binasch {\it et al.}, {\it Phys. Rev. B} {\bf 39}, 4828 (1989).
\bibitem{akhtar}M. Akhtar {\it et al.}, {\it npj 2D Mater. Appl.} {\bf 1}, 5 (2017).
\bibitem{Wang} V. Wang, Y.C. Liu, Y. Kawazoe, and W.T. Geng, {\it J. Phys. Chem. Lett.}, {\bf 6}, 4876 (2015);
P. Li and I. Appelbaum, {\it Phys. Rev. B}, {\bf 90}, 115439 (2014).
\bibitem{Rodin} A.S. Rodin, A. Carvalho, A.H. Castro Neto, {\it Phys. Rev. Lett.}, {\bf 112}, 176801 (2014)
\bibitem{LLi}L. Li {\it et al.}, {\it Nat. Nanotechnol.} {\bf 9}, 372 (2014).
\bibitem{yang}G. Yang {\it et al.}, {\it Phys. Rev. B} {\bf 94}, 075106 (2016).
\bibitem{kadioglu}Y. Kadioglu, {\it Phys. Rev. B} {\bf 96}, 245424 (2018).
\bibitem{mirzaei}M. Mirzaei {\it et al.}, {\it Phys. Rev. B} {\bf 98}, 045429 (2018). 
\bibitem{hu}T. Hu and J. Hong, {\it J. Phys. Chem. C} {\bf 119}, 8199 (2015).
\bibitem{jiang}J. Jiang {\it et al.}, {\it Comput. Mater. Sci.} {\bf 153}, 10 (2018).
\bibitem{luo}Y. Luo {\it et al.}, {\it Nanoscale Res. Lett.} {\bf 13}, 282 (2018).
\bibitem{babar}R. Babar and M. Kabir {\it J. Phys. Chem. C} {\bf 120}, 14991 (2016).
\bibitem{yu}W. Yu {\it et al.}, {\it Nano Express} {\bf 11}, 77 (2016).
\bibitem{wang} Y. Wang {\it et al.}, {\it Mater. Des.} {\bf 121}, 77 (2017).
\bibitem{li}J.Y. Li {\it et al.}, {\it Phys. Lett. A} {\bf 380}, 3928 (2016).
\bibitem{a}M. V. Kamalakar, B. N. Madhushankar, A. Dankert and S. P. Dash, 
{\it Small} {\bf 11}, 2209 (2015).
\bibitem{b}A. Avsar {\it et al.}, {\it Nat. Phys.} {\bf 13}, 888EP (2017).
\bibitem{c}P. Kumari {\it et al.}, {\it Physical Chemistry Chemical Physics} {\bf 22}, 5893 (2020). 
\bibitem{d}AK Nair {\it et al.}, {\it Physical Chemistry Chemical Physics} {\bf 21}, 23713 (2019).
\bibitem{dft}T.L. Loucks, {\it Augmented Plane Wave Method}
 (New York:Benjamin, 1967); O.K. Andersen, {\it Solid State Commun.} {\bf 13},
133 (1973); E. Wimmer, H. Krakauer, M. Weinert, and A.J. Freeman, {\it Phys. Rev. B} {\bf 24}, 864 (1981).
\bibitem{wien2k}P. Blaha, K. Schwarz, G.K.H. Madsen, D. Kvasnicka, and J. Luitz,
{\it WIEN2k, An Augmented Plane Wave + Local Orbitals Program for Calculating
Crystal Properties} (Wien:Karlheinz Schwarz, Techn. Universität Wien, (2001).
\bibitem{1404.5171} H. Lv {\it et al.}, {\it arXiv:1404.5171}.
\bibitem{nitro}L Shi, and X Luo, {\it J. Appl. Phys.} {\bf 125}, 233902 (2019). 
\bibitem{X-L-wang}X-Lin Wang, S Xue Dou, and C Znang {\it NPG Asia Materials} {\bf 2}, 31 (2010).
\bibitem{data}The data that support the findings of this study are available from the corresponding author upon reasonable request.
\end{thebibliography}
\end{document}